\documentclass[twocolumn,usenames,dvipsnames]{aastex701}
\usepackage{graphicx}
\usepackage{graphics}
\usepackage{amsmath}
\usepackage{amssymb}
\usepackage{url}
\usepackage{hyperref}
\usepackage{xcolor}
\usepackage{caption}

\newcommand{\OVI}{\ion{O}{6}}

\shorttitle{Circumgalactic \OVI\ Emission}
\shortauthors{Lochhaas et al.}

\begin{document}

\title{Figuring Out Gas \& Galaxies In Enzo (FOGGIE) XI: Circumgalactic \OVI\ Emission Traces Clumpy Inflowing Recycled Gas}

\author[0000-0003-1785-8022]{Cassandra Lochhaas}
\altaffiliation{NASA Hubble Fellow}
\affiliation{Center for Astrophysics $|$ Harvard \& Smithsonian \\
60 Garden St. \\
Cambridge, MA 02138, USA}
\email[show]{clochhaas@cfa.harvard.edu}

\author[0000-0003-1455-8788]{Molly S.\ Peeples}
\affiliation{Space Telescope Science Institute, 3700 San Martin Dr., Baltimore, MD 21218}
\affiliation{Center for Astrophysical Sciences, William H.\ Miller III Department of Physics \& Astronomy, Johns Hopkins University, 3400 N.\ Charles Street, Baltimore, MD 21218}
\email{molly@stsci.edu}

\author[0000-0002-2786-0348]{Brian W.\ O'Shea}
\affiliation{Department of Computational Mathematics, Science, \& Engineering, Michigan State University, 428 S. Shaw Lane, East Lansing, MI 48824}
\affiliation{Department of Physics \& Astronomy, 567 Wilson Road, Michigan State University, East Lansing, MI 48824}
\affiliation{Facility for Rare Isotope Beams, Michigan State University, 640 S. Shaw Lane, East Lansing, MI 48824}
\affiliation{Institute for Cyber-Enabled Research, 567 Wilson Road, Michigan State University, East Lansing, MI 48824}
\email{bwoshea@msu.edu}

\author[0000-0002-7982-412X]{Jason Tumlinson}
\affiliation{Space Telescope Science Institute, 3700 San Martin Dr., Baltimore, MD 21218}
\affiliation{Center for Astrophysical Sciences, William H.\ Miller III Department of Physics \& Astronomy, Johns Hopkins University, 3400 N.\ Charles Street, Baltimore, MD 21218}
\email{tumlinson@stsci.edu}

\author[0000-0002-0646-1540]{Lauren Corlies}
\affiliation{University of California Observatories/Lick Observatory, Mount Hamilton, CA 95140, USA}
\email{lcorlies@ucsc.edu}

\author[0009-0000-7559-7962]{Vida Saeedzadeh}
\affiliation{Center for Astrophysical Sciences, William H.\ Miller III Department of Physics \& Astronomy, Johns Hopkins University, 3400 N.\ Charles Street, Baltimore, MD 21218}
\email{vsaeedz1@jh.edu}

\author[0000-0001-9158-0829]{Nicolas Lehner}
\affiliation{Department of Physics and Astronomy, University of Notre Dame, Notre Dame, IN 46556}
\email{nlehner@nd.edu}

\author[0000-0002-1685-5818]{Anna C.\ Wright}
\affiliation{Center for Computational Astrophysics, Flatiron Institute, 162 Fifth Avenue, New York, NY 10010}
\email{awright@flatironinstitute.org}

\author[0000-0002-0355-0134]{Jessica K.\ Werk}
\affil{University of Washington, Department of Astronomy, Seattle, WA 98195, USA}
\email{jwerk@uw.edu}

\author[0000-0001-7813-0268]{Cameron W.\ Trapp}
\affiliation{Center for Astrophysical Sciences, William H.\ Miller III Department of Physics \& Astronomy, Johns Hopkins University, 3400 N.\ Charles Street, Baltimore, MD 21218}
\email{ctrapp2@jhu.edu}

\author[0000-0001-7472-3824]{Ramona Augustin}
\affiliation{Leibniz-Institut f{\"u}r Astrophysik Potsdam (AIP), An der Sternwarte 16, 14482 Potsdam, Germany}
\email{raugustin@aip.de}

\author[0000-0003-4804-7142]{Ayan Acharyya}
\affiliation{INAF - Astronomical Observatory of Padova, vicolo dell’Osservatorio 5, IT-35122 Padova, Italy}
\email{ayan.acharyya@inaf.it}

\author[0000-0002-6804-630X]{Britton D.\ Smith}
\affiliation{Institute for Astronomy, University of Edinburgh, Royal Observatory, EH9 3HJ, UK}
\email{Britton.Smith@ed.ac.uk}

\author[0000-0001-7936-0831]{Carlos J.\ Vargas}
\affiliation{Steward Observatory, University of Arizona, 933 N Cherry Avenue, Tucson, AZ 85721, USA}
\email{cjvargas@arizona.edu}

\begin{abstract}

The circumgalactic medium (CGM) is host to gas flows into and out of galaxies and regulates galaxy growth, but the multiphase, diffuse gas in this region is challenging to observe. We investigate the properties of gas giving rise to \OVI\ emission from the CGM that upcoming missions, such as the \emph{Aspera} SmallSat, will be able to map in local galaxies. We use the FOGGIE simulations to predict the \OVI\ emission from edge-on galaxies across the redshift range $z=1\rightarrow0$. \OVI\ emission is brightest surrounding small, clumpy structures near the galaxy where the gas density is high. Most of the \OVI\ surface brightness originates from collisionally ionized, $T\sim10^{5.5}$ K, inflowing gas and is not preferentially aligned with the major or minor axis of the galaxy disk. Simulated galaxies with higher halo masses, higher median CGM gas density, and higher star formation rates produce brighter and more widespread \OVI\ emission in their CGM. We show that while \OVI\ emission primarily originates in inflowing gas, turning off outflows in a simulation without star formation feedback eliminates most of the \OVI\ emission. Enrichment from feedback is necessary to mix with the inflowing gas and allow it to glow in \OVI. Collectively, our findings point towards a picture where \OVI\ emission traces warm, ionized envelopes of cooler clouds that are accreting onto the galaxy in a metal-enriched galactic fountain. Finally, we show that the detection limit of \emph{Aspera} is sufficient to detect \OVI\ emission tens of kpc from the galaxy center for $\sim L^\star$ galaxies.

\end{abstract}


\section{Introduction} \label{sec:intro}

The circumgalactic medium (CGM) is the diffuse gas surrounding galaxies that acts as the gateway between galaxies and the Universe. Gas flowing inward from the intergalactic medium (IGM) passes through the CGM before it reaches galaxies, and gas launched out from galaxies also enters into the CGM. As such, the CGM sits in a unique position as both promoter and regulator of galaxy growth, but this diffuse gas reservoir is challenging to observe. Simulations help us understand the properties and dynamics of CGM gas, but we must determine how these properties manifest in observables.

The CGM is multiphase, spanning a $\gtrsim4$ dex spread in both gas density and temperature \citep{Tumlinson2017,Faucher2023}. The most common way to observe the $T\sim10^{4-6}$ K gas has been through \ion{H}{1} and metal line absorption towards background light sources, such as bright quasars. Several ionization states of prominent metals have absorption transitions in the rest-frame ultraviolet (UV) or optical, from low-ions like \ion{Mg}{2}, \ion{C}{2}, and \ion{Si}{2} that are expected to trace cooler phases to intermediate and high ions like \ion{C}{4}, \ion{Si}{4}, and \OVI\ that trace warmer gas. Surveys that search for absorption in the CGM of nearby galaxies ubiquitously find high covering fractions of gas within $\sim 200$ kpc of galactic disks \citep[e.g.,][]{Chen2009,Keeney2017,Rudie2019, Prochaska2011,Werk2014,Johnson2015}.

In particular, the $\lambda\lambda1031, 1037$ doublet of \OVI\ has a high oscillator strength, and oxygen is an abundant element, so absorption is found often. \citet{Chen2009} and \citet{Tumlinson2011} found that star-forming galaxies are more likely to host \OVI\ absorption in their halos than passive galaxies. \citet{Oppenheimer2016} argued that this dichotomy could be driven by halo mass: more massive galaxies are both more likely to be quenched and have a virial temperature higher than the temperature at which \OVI\ ionization fractions peak. \citet{Dutta2025} found that low-mass galaxies do not host as much \OVI\ in their CGM, suggesting that \OVI\ absorption peaks around $L_\star$ galaxies. \citet{Tchernyshyov2023} analyzed the incidence of \OVI\ absorption surrounding galaxies in narrow bins of stellar mass and found that even at a single mass, galaxies with higher SFR hosted more \OVI\ absorption in their CGM. This is consistent with the findings of \citet{Kacprzak2015} that \OVI\ absorption is preferentially located along both the minor and major axes of galaxies, and the findings of \citet{Kacprzak2025} that 60\% of \OVI\ absorbers are co-rotating with the galaxy disk, suggesting \OVI\ traces complex kinematics of both inflows that fuel star-forming galaxies and outflows that are launched by stellar feedback.

Absorption-line observations provide limited spatial information: there are typically only $\sim$1--2 bright quasars behind any given galaxy's CGM \citep[nearby galaxies like M31 and the LMC are exceptions,][]{Lehner2015,Lehner2020,Krishnarao2022,Lehner2025}, and it is difficult if not impossible to determine where along the line of sight the absorbing gas is located within the galaxy system. In addition, \OVI\ can be produced through either photoionization or collisional ionization, in ionization equilibrium or in non-equilibrium cooling flows \citep{Gnat2007,Oppenheimer2013a}. Gas must be diffuse for \OVI\ to be photoionized by the extragalactic UV background (EUVB), in which case it traces gas at $T\sim10^4$ K \citep{Stern2018,Taira2025}. If the gas is warmer and/or denser, \OVI\ is instead collisionally ionized and peaks at $T\sim10^{5.5}$ K \citep{Gnat2007}, which is a ``transition temperature" regime that is efficient at radiatively cooling and is thus not expected to have a long lifetime \citep{Wakker2012,McQuinn2018}. For all these reasons, it is difficult to determine the properties and physics of the gas traced by \OVI, despite its observed prevalence.

Observing the CGM in emission provides significantly more information than in absorption. Emission maps give 2D projected spatial information of the density of emitting ions and line-of-sight velocity can also be inferred from emission lines if the spectral resolution of the instrument is sufficient. Studies that use the Multi Unit Spectrographic Explorer \citep[MUSE,][]{Bacon2010} or the Keck Cosmic Web Imager \citep[KCWI,][]{Morrissey2018} to image the CGM in optical lines, such as [\ion{O}{2}] or [\ion{O}{3}], find extended halos of ionized gas surrounding starburst galaxies \citep{Nielsen2024} and intermediate-redshift quasar hosts \citep{Chen2023}. A small handful of studies with both the \emph{Far Ultraviolet Spectroscopic Explorer} \citep[FUSE,][]{Moos2000} and the \emph{Hubble Space Telescope} have detected \OVI\ emission from gas near galaxy disks \citep{Otte2003,Hayes2016,Chung2021a} and in strong outflows \citep{Kim2024,Ha2025}, but the limited sensitivity and field of view of these instruments means there has not yet been a true ``map" of \OVI\ emission from the CGM.

Several new and upcoming CGM emission instruments are beginning to image the gas surrounding nearby galaxies, providing significantly more information than was previously possible with absorption line observations. The Circumgalactic H$\alpha$ Spectrograph \citep[CHaS,][]{Melso2024} and the Dragonfly Spectral Line Mapper \citep{Lokhorst2022} have recently detected H$\alpha$ and [\ion{N}{2}] emission in the CGM of NGC 1068 and the M 81 group. \emph{Aspera} is a SmallSat mission that will be launched in 2026 to image the CGM of a handful of nearby edge-on galaxies in \OVI\ \citep{Chung2021b}. In this paper, we use a detection limit similar to that for \emph{Aspera} to predict the \OVI\ emission we expect to be detectable. JUNIPER \citep{Witt2025} is a proposed CubeSat instrument that will target multiple UV metal lines, including \OVI, in the CGM of nearby galaxies. By obtaining the spatial distribution and brightness of \OVI-emitting gas, we will soon have a better understanding of the properties of gas that hosts \OVI: is it volume-filling, organized into clumps or filaments, or preferentially oriented in some way with respect to the host galaxy? Previous predictions of \OVI\ emission from simulations \citep{Corlies2016,Augustin2019,Corlies2020} suggest it may not be homogeneous and could pick out gaseous structures in particular ranges of CGM temperature and density.

With new observations on the horizon, it is crucial to link the underlying gaseous properties to the observable emitting gas. In this paper, we predict the \OVI\ emission from the CGM surrounding galaxies oriented edge-on in the Figuring Out Gas \& Galaxies In Enzo (FOGGIE) suite of simulations, investigate the origins of \OVI-emitting gas, and search for correlations between circumgalactic \OVI\ surface brightness and properties of the host galaxy. \citet{Saeedzadeh2025} uses the FOGGIE simulations to predict emission in a range of optical and UV lines, focusing on the instrumental effects of sensitivity limit and spatial and kinematic resolution. Here, we focus on only \OVI\ to understand the physical properties of gas giving rise to emission and defer specific instrumental predictions to \citet{Saeedzadeh2025}.

Section \ref{sec:methods} describes the FOGGIE simulations (\ref{subsec:foggie}) and the process for making \OVI\ emission predictions (\ref{subsec:emis_predict}). Section \ref{sec:results} presents the main results of the morphology of \OVI\ emission (\ref{subsec:morphology}), the properties of the gas that primarily contributes to \OVI\ emission (\ref{subsec:emiss_props}), and trends of \OVI\ emission with halo properties (\ref{subsec:halo_prop}) and stellar feedback (\ref{subsec:feedback}). In Section \ref{sec:discussion}, we summarize our uncovered picture of \OVI\ emission as a tracer of the baryon cycle (\ref{subsec:baryon_cycle}), emphasize that our emission predictions are lower limits (\ref{subsec:lower_limit}), compare our findings to other simulated emission predictions (\ref{subsec:comparison}), discuss the implications of this study for the upcoming \emph{Aspera} mission and other future CGM emission studies (\ref{subsec:Aspera}), and present some caveats of the simulations and methodology (\ref{subsec:caveats}). We conclude in Section \ref{sec:conclusions}.

%%%%%%%%%%%%%%%%%%%%%%%%%%%%%%%%%%%%%%%%%%%%%%%%%%%%%%%%%%%%%%%%%%%%%%%%%%%%%%%%%%%%%
\section{Methods} \label{sec:methods}

In this section, we give an overview of the FOGGIE simulations (Section~\ref{subsec:foggie}) and describe the methods by which we post-process the simulations to obtain \OVI\ emissivities and surface brightnesses (Section~\ref{subsec:emis_predict}).

\subsection{FOGGIE Simulations} \label{subsec:foggie}

The Figuring Out Gas \& Galaxies In Enzo (FOGGIE) simulations have been introduced and described in \citep{Peeples2019,Wright2024}, but here we give a brief overview of the important components of the simulations for this work. FOGGIE is a suite of cosmological zoom-in simulations run with the adaptive mesh refinement hydrodynamics code Enzo \citep{Bryan2014,BrummelSmith2019}. We analyze all six galaxies in the FOGGIE suite. They are chosen from a $\Lambda$CDM $100 h^{-1}$ Mpc cosmological volume with $\Omega_m = 0.285$, $\Omega_{\Lambda} = 0.715$, $\Omega_{b} = 0.0461$, $H_0 =  69.5$ km s$^{-1} \rm Mpc^{-1}$ and $\sigma _8 = 0.82$ \citeauthor{Planck2016} \citeyear{Planck2016}). All six galaxies have roughly Milky Way mass ($\sim10^{12}$M$_{\odot}$) dark matter halos by $z=0$ and no major mergers with a mass ratio greater than 1:10 below $z=2$ (with the exception of the halo Squall, which has a 1:2 merger at $z=0.7$). Star formation follows the scheme of \citet{Cen1992}, where star particles form in dense cells with a minimum mass of 1000 M$_{\odot}$ at $z>2$, ramping up to $10^4$ M$_\odot$ by $z\leq1$. Stellar feedback is implemented by injecting thermal energy into the cells immediately surrounding star particles at the rate of $10^{51}$ ergs per $100$ M$_{\odot}$ of stellar mass formed, spread over several dynamical times of the star particle. Feedback deposits 25\% of the star particle mass and 2.5\% of the star particle mass as metals \citep[for full details of star formation and feedback scheme, see][]{Wright2024}. Radiative cooling is implemented using the Grackle chemistry and cooling library \citep{Smith2017}, assuming a \citet{Haardt2012} extragalactic UV background with a self-shielding approximation and solar relative metal abundances. 

The FOGGIE galaxies have larger stellar masses than would be expected from their halo masses and the stellar-mass-to-halo-mass relation \citep[see][]{Wright2024}, most likely owing to the adopted feedback scheme. The feedback is purely thermal, which means that the galactic outflows in FOGGIE are typically very fast and hot, with $v_\mathrm{outflow}\approx 2000$ km s$^{-1}$ and $T_\mathrm{outflow}\approx 10^7$ K. The outflows are usually also quite diffuse and do not carry much mass. At $z=0$, the outflowing gas mass within the virial radius (defined with $v_\mathrm{outflow}>200$ km s$^{-1}$, approximately the escape velocity for these halos) is $\approx10^{8\pm0.5}M_\odot$ across the six halos. In Section~\ref{subsec:emiss_props}, we show that such a hot form of feedback means there is very little \OVI\ emission originating in outflows.

Table~\ref{tab:halo_props} lists properties of the six FOGGIE galaxies used in this work at $z=0$. The virial radius $R_\mathrm{vir}$ is calculated as the radius within which the total mass density (of dark matter, stars, and gas) is $200\times$ the critical density of the Universe. The halo mass listed in Table~\ref{tab:halo_props}, $M_h$, is this total mass within $R_\mathrm{vir}$. The star formation rate (SFR) is calculated as the mass of star particles formed within 20 kpc of the center of the galaxy (determined by the location of the peak of the dark matter distribution) within the previous 10 Myr, divided by 10 Myr. The CGM fraction $f_\mathrm{CGM}$ is calculated as the mass of gas outside of the disk \citep[defined using the gas density cut of][]{Lochhaas2023}, divided by the halo mass $M_h$, and normalized to the average cosmic baryon fraction $\Omega_b/\Omega_m\approx0.16$ \citep{Crain2007}.

\begin{table}[]
    \centering
    \begin{tabular}{c|c c c c}
         & $M_h$ & $R_\mathrm{vir}$ & SFR & $f_\mathrm{CGM}$ \\ 
         &  (M$_\odot$) & (kpc) & (M$_\odot$ yr$^{-1}$) & \\ \hline
        Tempest & $5.1\times10^{11}$ & 168 & 1.3 & 0.075 \\
        Squall & $8.0\times10^{11}$ & 196 & 2.1 & 0.094 \\
        Maelstrom & $1.0\times10^{12}$ & 212 & 2.0 & 0.094 \\
        Blizzard & $1.1\times10^{12}$ & 220 & 2.1 & 0.010\\
        Hurricane & $1.7\times10^{12}$ & 252 & 4.7 & 0.094 \\
        Cyclone & $1.7\times10^{12}$ & 253 & 13.5 & 0.113
    \end{tabular}
    \caption{The halo mass $M_h$, virial radius $R_\mathrm{vir}$, star formation rate SFR, and CGM fraction $f_\mathrm{CGM}$ (normalized to an average halo baryon fraction of $0.16$) of each of the six FOGGIE galaxies analyzed in this paper at $z=0$.}
    \label{tab:halo_props}
\end{table}

The FOGGIE simulations do not implement feedback from active galactic nuclei (AGN), which other large cosmological simulations have shown is an important source of feedback for halos with $M_h\gtrsim10^{12}M_\odot$ \citep{Crain2023}, particularly in its ability to evacuate baryons from halos. The CGM fraction of the FOGGIE galaxies ranges from 7.5\% to 11\%, lower than the CGM fractions of SIMBA, IllustrisTNG, EAGLE, and Romulus \citep{Crain2023} at $M_h\sim 10^{11}$--$10^{12}M_\odot$, all of which include prescriptions for AGN feedback. This is likely due to the fact that the FOGGIE galaxies form too many stars, as discussed above, rather than ``evacuation" of the halo baryons by feedback, but this indicates the lack of AGN feedback is not creating an overly-massive CGM that could be affecting \OVI\ emission results. AGN feedback may also affect the thermodynamic properties of the CGM, which may have an impact on \OVI\ emission predictions, and we discuss this further in Section~\ref{subsec:caveats}.

In addition to the adaptive mesh refinement that refines the gas grid where the gas density is large, FOGGIE also implements a forced refinement region and cooling refinement \citep[first introduced in][]{Simons2020}. Forced refinement sets an upper bound on the size of each gas cell within a box that tracks the galaxy through its evolution. The forced refinement region is a cube 288 comoving kpc (ckpc) on a side, centered on the galaxy halo. Within this region, the maximum size of any gas cell is 1.1 ckpc, but cells can be smaller in locations where other refinement criteria act. Cooling refinement acts similarly to the adaptive refinement, but instead of refining on gas density, it refines on the cooling length of the gas, $l_\mathrm{cool}\sim t_\mathrm{cool}c_s$, where $t_\mathrm{cool}$ is the cooling time and $c_s$ is the sound speed of the gas. The cooling refinement operates only within the forced refinement region. The maximum allowed refinement produces cell sizes of 274 comoving pc (cpc). Typically, a combination of density and cooling refinement ensures the galaxy and its extended gaseous disk are refined to the highest degree, while the larger CGM is refined to the forced refinement level of 1.1 ckpc cells. Forced high resolution in the CGM of simulated galaxies tends to produce a higher degree of cold gas structure and more cold gas mass in the CGM \citep{Hummels2019,Suresh2019,vandeVoort2019}.

\subsection{Making Emission Predictions} \label{subsec:emis_predict}

We calculate the emissivity and surface brightness of \OVI\ following the process of \citet{Corlies2020} and \citet{Saeedzadeh2025}. We use CIAOLoop\footnote{\url{https://github.com/brittonsmith/cloudy_cooling_tools.git}} \citep{Smith2009} to generate a grid of CLOUDY \citep{Ferland2013,Chatzikos2023} tables to obtain the emissivity of \OVI\ as a 2D function of gas temperature and density, assuming solar metallicity and solar relative abundances. The CLOUDY tables assume both collisional ionization and photoionization equilibrium, using the same \citep{Haardt2012} extragalactic UV background\footnote{\citet{Taira2025} found that the \citet{Haardt2012} UV background can underestimate the number and column density of \OVI\ absorbers in the CGM of the FOGGIE galaxies at $z=2.5$ by up to $\sim0.5$ dex, so this may also be true at the $z\sim0$ redshifts explored in this work. However, we show in Section~\ref{subsec:emiss_props} that \OVI\ emission appears to primarily trace collisionally-ionized, rather than photoionized, gas, so the impact of the photoionizing UV background on the emission is likely small.} with which the FOGGIE simulations were run as the photoionizing source. We use these tables in post-processing as lookup tables to determine the emissivity of each gas cell in the simulation from its hydrogen number density and temperature. We then linearly scale the emissivity, which is calculated assuming solar metallicity, to the simulated metallicity of the gas cell. The simulation has a single metallicity field and does not track individual metal elements, so we assume solar relative elemental abundances throughout.

\begin{figure*}
    \centering
    \includegraphics[width=\linewidth]{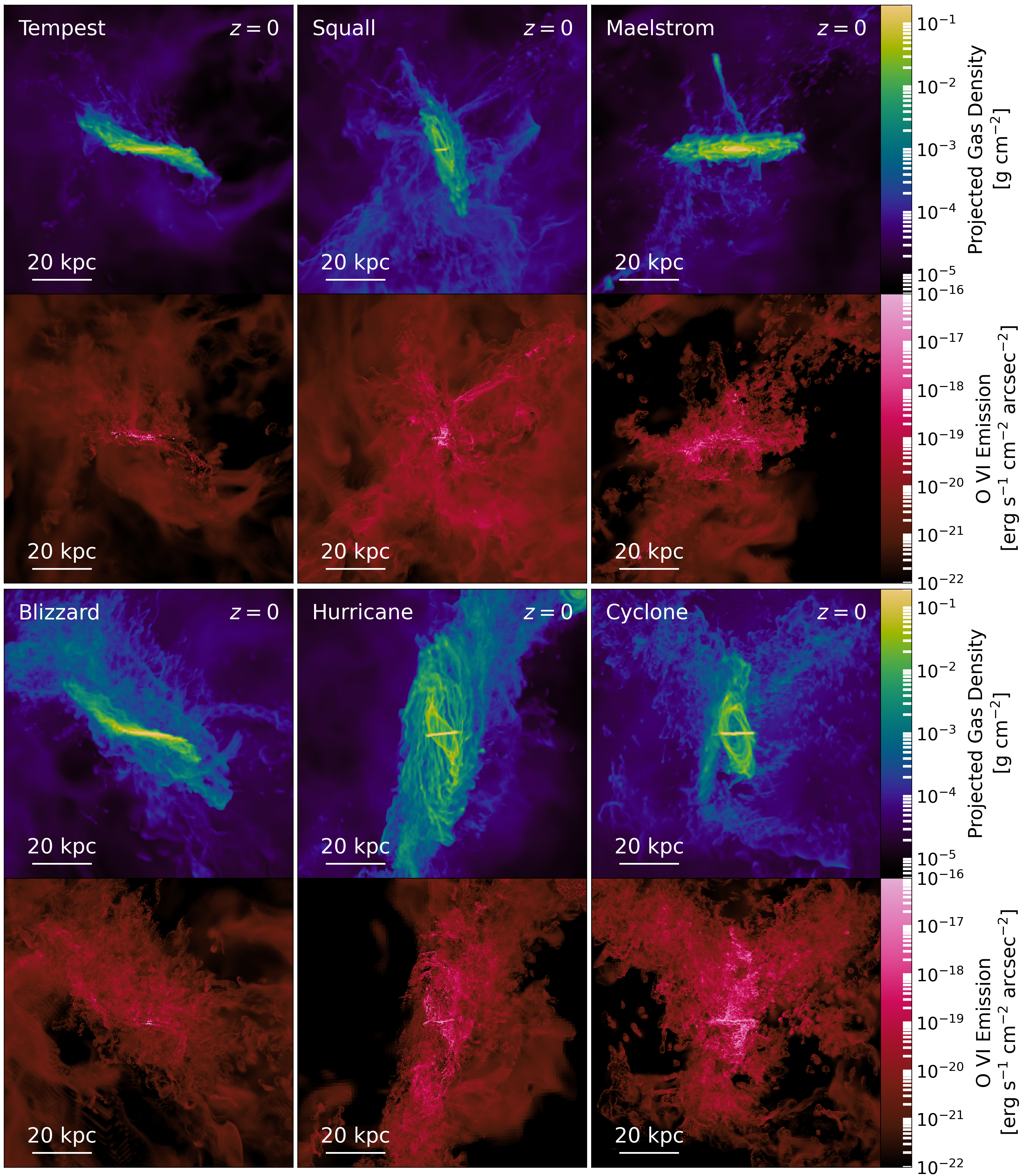}
    \caption{Projected gas density (top of each subpanel) and surface brightness of \OVI\ emission (bottom of each subpanel) from the six galaxies in the FOGGIE suite, each oriented edge-on and shown at $z=0$. The morphology of the emission is highly variable from one galaxy to the next, and shows both large, smooth structures and small, clumpy structures.} \label{fig:maps_z0}
\end{figure*}

Because we calculate \OVI\ emission in post-processing without information about the gas's time evolution, we make the assumption that \OVI\ is in both photo- and collisional ionization equilibrium. Many works have found that in certain conditions, such as when gas is rapidly expanding \citep{Gray2019b}, undergoing shocks \citep{Bordoloi2017}, or actively radiatively cooling \citep{Gnat2007}, \OVI\ can be over- or under-ionized relative to ionization equilibrium. Without time-dependent radiative transfer calculations as the simulation runs, we cannot capture the effects of non-equilibrium ionization or chemistry \citep[see][for the impacts of time-dependent radiative transfer on the CGM for high-redshift galaxies]{Cadiou2025}. However, the small and dense structures where \OVI\ emission is predominantly found may be less susceptible to non-equilibrium ionization conditions commonly found in shocks or rapid gas expansion, so non-equilibrium ionization may not strongly affect the brightest regions of \OVI\ emission. We explicitly test the validity of the ionization equilibrium assumption in Section~\ref{subsec:emiss_props}.

We also assume that there are no local sources of photoionization and that all ionizing photons come from the extragalactic UV background. Starlight from the galaxy could be contributing additional ionizing photons in the inner regions of the CGM, where we find the brightest \OVI\ emission. The ionizing photons from starlight are not energetic enough to affect the predictions of such a highly ionized ion as \OVI, but local ionization is important for lower ions \citep{Katz2022}. \citet{Nielsen2024} examined the ionization mechanisms of [\ion{O}{2}], [\ion{O}{3}], and H$\beta$ emission in the disk and inner CGM of a low-redshift galaxy. They found the ionization mechanism of the emitting gas (as inferred by line ratios) shifted from \ion{H}{2} region ionization within the disk, to shock-dominated ionization just outside the disk up to $\sim6$ kpc, to extragalactic UV background ionization at distances $\gtrsim6$ kpc, suggesting that the assumption of UV background ionization equilibrium may not be accurate close to the galaxy. Because we focus on \OVI\ emission predictions here, we do not expect the lack of local photoionization sources to affect our results.

We convert the volumetric emissivities reported by CLOUDY into surface brightness by projecting the simulated galaxy from 3D to 2D and integrating the emissivity along the line of sight. We choose to use ergs s$^{-1}$ cm$^{-2}$ arcsec$^{-2}$ as the units of surface brightness in this paper, but see Appendix A of \citet{Saeedzadeh2025} for unit conversions. Throughout this paper, we choose to view the simulated galaxies edge-on, to minimize confusing emission from the galaxy disk with that from the CGM \citep[for a broader range of views see][]{Saeedzadeh2025}. We calculate the angular momentum vector of the disk as the total angular momentum of stars that are $<10$ Myr old and within 15 kpc from the center of the galaxy (as determined by the peak of the dark matter density distribution), then orient the edge-on emission maps such that the angular momentum vector points north. Note that several of the FOGGIE galaxies have warped disks or polar rings that may make this calculation imprecise, leading to slightly inclined disks in the emission maps \citep{Simons2025,Trapp2025}. We generate emission maps within $\pm50$ kpc from the center of the galaxy with an image resolution of $370\times370$ pixels (pixel side length $\approx0.27$ kpc, the same as the highest resolution gas cell in the simulation).

%%%%%%%%%%%%%%%%%%%%%%%%%%%%%%%%%%%%%%%%%%%%%%%%%%%%%%%%%%%%%%%%%%%%%%%%%%%%%%%%%%%%%
\section{Results} \label{sec:results}

\subsection{\OVI\ Emission Morphology} \label{subsec:morphology}

\begin{figure}
    \centering
    \includegraphics[width=\linewidth]{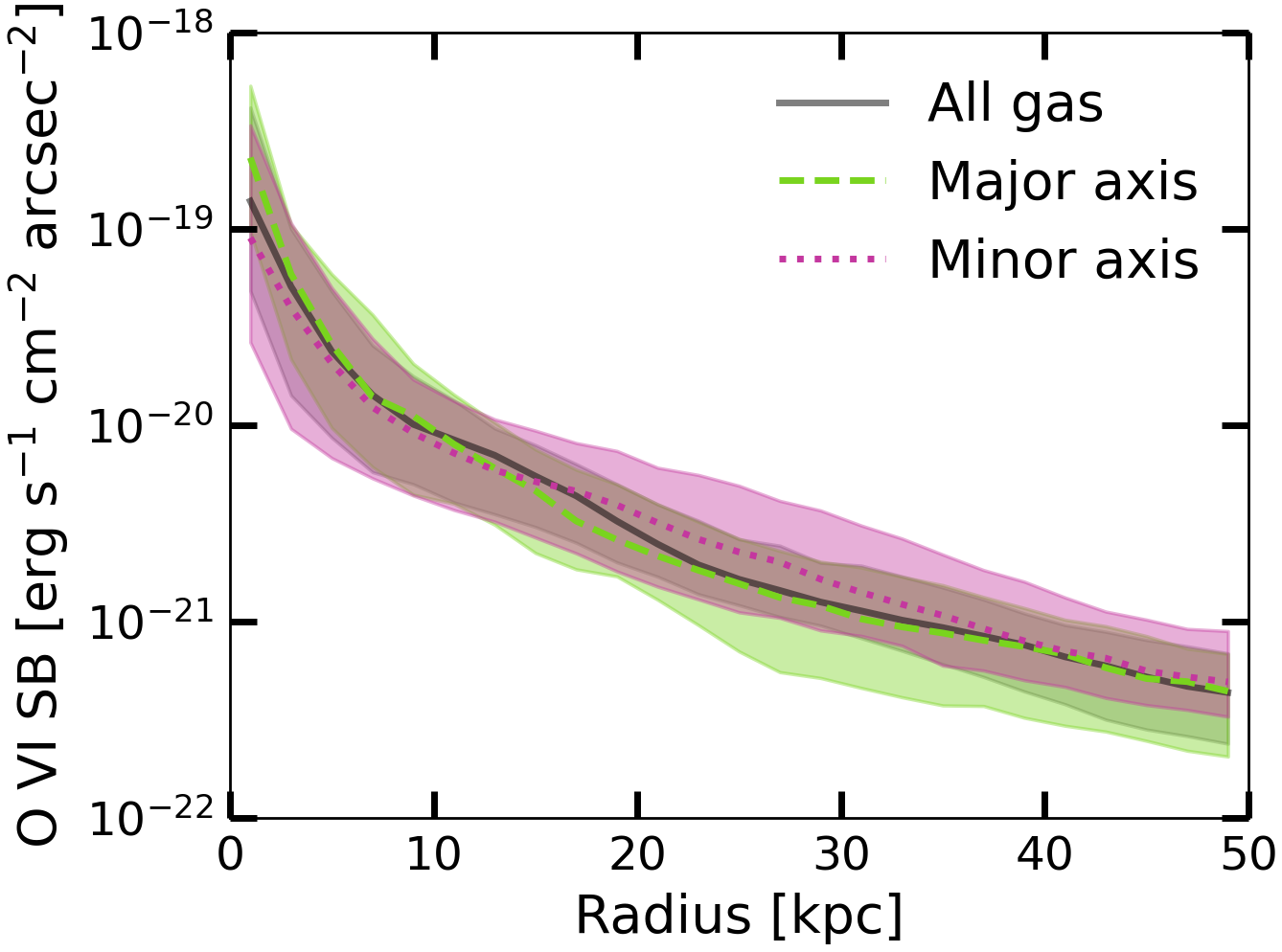}
    \caption{The average of the surface brightness profiles of all six galaxies' \OVI\ emission as a function of galactocentric distance (black curve) compared to the \OVI\ SB profile of only that gas along the major axis (green dashed curve) or along the minor axis (pink dotted curve), averaged over $z=0-0.1$. Shaded regions show the standard deviation of the SB profile among the six galaxies and over time. There is no strong difference in \OVI\ SB profile between the major axis and the minor axis, suggesting no strongly preferred orientation for \OVI-emitting gas.} \label{fig:sb_axes_z0}
\end{figure}

Figure~\ref{fig:maps_z0} shows maps of projected gas density (left column) and \OVI\ surface brightness (right column) for the six FOGGIE galaxies at $z=0$, oriented edge-on. In each case, the galaxy disk is located in the center of the image, with the direction of the disk's angular momentum vector pointing north. The galaxies show significant differences from one another in both their gas density and  \OVI\ emission maps. In several cases, there is an extended gaseous disk oriented at an inclined angle from the smaller central disk. For Tempest and Blizzard, the extended disk shows an S-warp, and for Squall, Hurricane, and Cyclone the extended disk appears nearly polar relative to the inner disk. The morphology of extended gas disks in the FOGGIE galaxies is further explored in \citet{Trapp2025}, who show these features are created by mergers and misaligned gas accretion and can persist for several Gyr.

Here, we focus on how this gas structure translates into \OVI\ emission maps: there are clear structures of high surface brightness that appear to roughly align with the denser gas regions visible in the left panels, but there does not appear to be much pattern to the location of the brightest structures relative to the edge-on galaxy disk itself, reflecting the variety of structures in the CGM of the FOGGIE galaxies. The brightest \OVI\ emission appears closer to the galaxy disks, and is generally structured, with significant variation in brightness on small scales. Each galaxy halo also shows \OVI\ emission from extended, smoother structures, particularly further from the galaxy. By comparing to the left panels, it is clear these smooth \OVI\ emission regions originate from very low-density gas, while the clumpier, brighter \OVI\ emission regions seem to form shells at the edges of high-density gas. The smooth structures also tend to occupy more of the map area, and out to larger distances from the galaxy, than the clumpy structures that are located primarily within $\sim30$ kpc of the galaxy's center. \citet{Augustin2025} found that dense CGM clouds tend to be found closer to the FOGGIE galaxies at $z=1$, and by eye we can see a similar trend for the \OVI-traced clumps at $z=0$ as well.

\begin{figure}
    \centering
    \includegraphics[width=\linewidth]{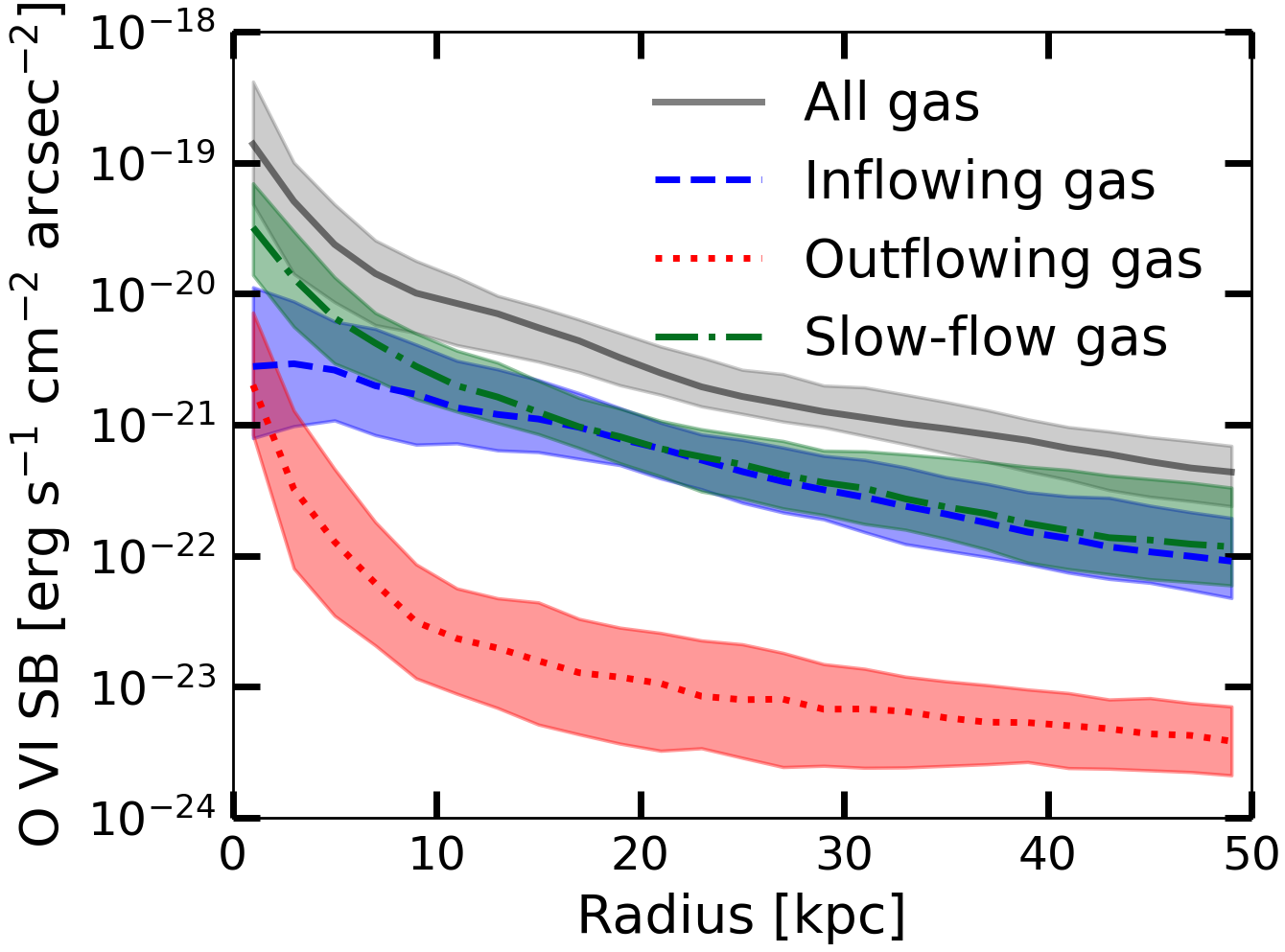}
    \caption{The average of the surface brightness profiles of all six galaxies' \OVI\ emission as a function of galactocentric distance (black curve) compared to the \OVI\ SB profile of only that gas that is radially inflowing with $v_r<-100$ km s$^{-1}$ (blue dashed curve), radially outflowing with $v_r>100$ km s$^{-1}$ (red dotted curve), or ``slow-flow" gas with $-100 < v_r < 100$ km s$^{-1}$ (green dot-dashed curve) averaged over $z=0-0.1$. The black curve is the same as in Figure~\ref{fig:sb_axes_z0}. Shaded regions show the standard deviation of the SB profiles among the six simulated galaxies and over time. The \OVI\ surface brightness originates roughly equally from inflowing gas and slow-flow gas, with only a minor contribution from outflowing gas.}
    \label{fig:sb_flows_z0}
\end{figure}

\begin{figure}
    \centering
    \includegraphics[width=\linewidth]{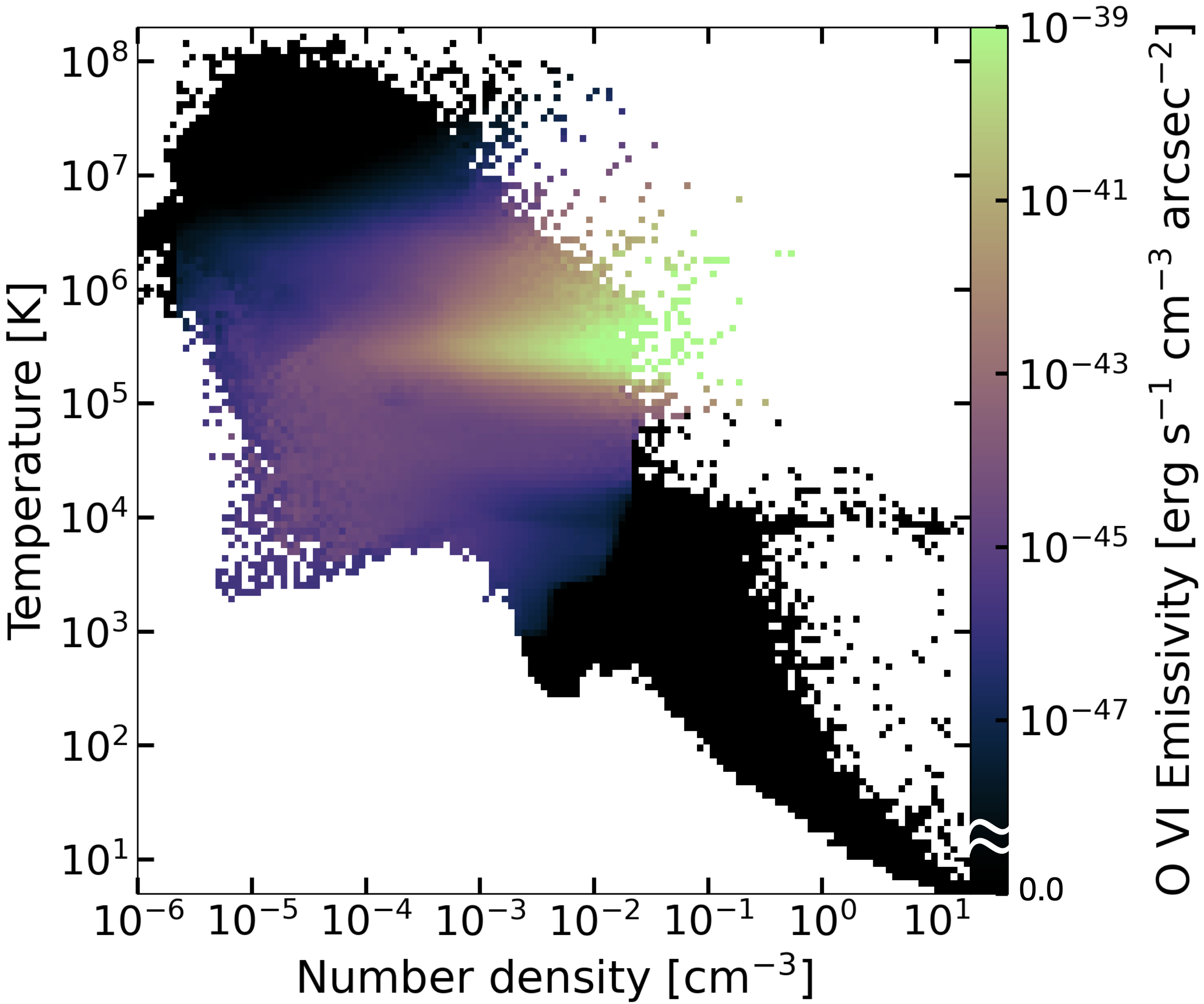}
    \caption{A temperature-density phase plot of the Maelstrom CGM at $z=0$, color-coded by mass-weighted average \OVI\ emissivity (black colors indicate where there is gas mass but no \OVI\ emission). While there is some emission from cooler gas phases near $T\sim10^4$, the majority of the emission comes from $T\sim10^5$--$10^6$ K gas.} \label{fig:phase}
\end{figure}

To investigate if \OVI\ emission has a preferred orientation surrounding edge-on galaxies, in Figure~\ref{fig:sb_axes_z0} we plot the \OVI\ surface brightness (SB) profile as a function of galactocentric radius (black curve) as well as the SB profiles for only that gas along the major axis (green dashed) or the minor axis (pink dotted). The gas along the major axis is defined as all gas within $\pm45^\circ$ of a horizontal line at $y=0$, and gas along the minor axis is similarly defined as all gas within $\pm45^\circ$ of a vertical line at $x=0$ in the image coordinate systems of Fig.~\ref{fig:maps_z0}. The profiles are calculated as the median SB within each spherical annulus (of radial width 2 kpc) for each halo, within the major and minor axes definitions. We calculate the SB profiles of 50 time snapshots spaced evenly between $z=0.1$ and $z=0$ (roughly every $\sim26.5$ Myr) for each halo, then average across all these snapshots and all six halos. The shaded region in Fig.~\ref{fig:sb_axes_z0} shows the standard deviation of the SB profiles across these time snapshots and the six galaxies. The SB profiles along the major axis and minor axis are not significantly different from one another, suggesting that \OVI\ emission has no preferred orientation surrounding the FOGGIE galaxies. This is consistent with the apparent randomness of the bright \OVI\ structures in Fig.~\ref{fig:maps_z0} relative to the galaxy disk orientation. We perform this averaging over time and across the six FOGGIE galaxies to determine if there are any general trends in \OVI\ emission orientation that are more difficult to see on a halo-by-halo basis. Most of the individual galaxies at $z=0$ also have consistent \OVI\ SB profiles between their major and minor axes, with the two exceptions of Maelstrom and Hurricane: Maelstrom has a brighter SB profile along the minor axis than along the major axis beyond $r = 20$ kpc (they are the same within $r = 20$ kpc), and Hurricane's minor axis SB profile dominates over the major axis profile at all radii, due to its extended polar disk generating \OVI\ emission primarily along the minor axis.

\subsection{What Gas Contributes to \OVI\ Emission?} \label{subsec:emiss_props}

\begin{figure*}
    \centering
    \includegraphics[width=\textwidth]{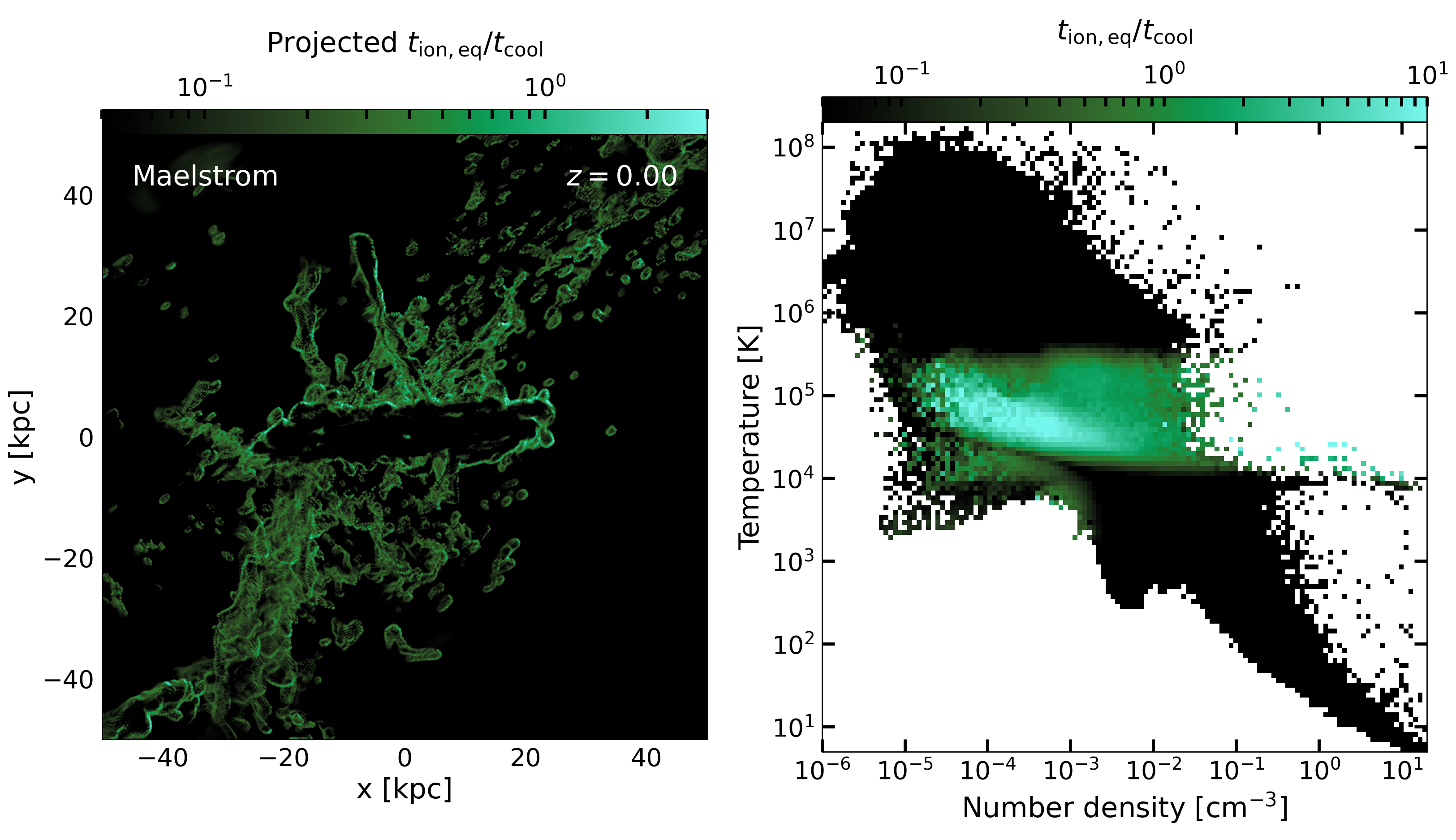}
    \caption{Left: A projection of the ratio of \OVI\ ionization equilibration time to cooling time, $t_\mathrm{ion,eq}/t_\mathrm{cool}$, for the Maelstrom halo at $z=0$. Right: A phase plot of gas within 50 kpc of the center of Maelstrom at $z=0$, color-coded by $t_\mathrm{ion,eq}/t_\mathrm{cool}$. Wherever $t_\mathrm{ion,eq}/t_\mathrm{cool}\gtrsim1$, the assumption of ionization equilibrium breaks down. This predominantly occurs on the edges of clouds and the edge of the extended gas disk, in interface regions of $T\sim10^{4.5}$--$10^5$ K where cooling times are short.}
    \label{fig:tcool_teq}
\end{figure*}

While we have determined that the \OVI\ emission reveals both clumpy and smooth structures, and that the clump shells are much brighter, it is difficult to determine from the projected images in Fig.~\ref{fig:maps_z0} what role these structures play in the baryon cycle or the evolution of the galaxy. Using the 3D information available in the simulations, we can determine if the \OVI\ emission traces inflowing, accreting gas fueling the galaxy or outflowing, ejected gas launched by feedback from star formation. This information is not available in observed emission maps, but by using the simulations to understand the typical properties of gas contributing to \OVI\ emission, we can infer what physical processes upcoming \OVI\ emission maps are likely to be probing.

Figure~\ref{fig:sb_flows_z0} shows the SB profile of \OVI\ emission from all gas (black curve), from radially inflowing gas with velocities faster than 100 km s$^{-1}$ (blue dashed curve), from radially outflowing gas with velocities faster than 100 km s$^{-1}$ (red dotted curve), and from gas that is not participating in either of these strong flows, which we term ``slow-flow" gas. The inflowing velocity cut of $-100$ km s$^{-1}$ captures gas within at least a third of the free-fall velocity for these galaxies within 50 kpc from the galaxy centers, while the outflowing velocity cut of 100 km s$^{-1}$ captures gas within $\sim0.5\times$ the escape velocity. These are both conservative cuts to ensure that the gas in these regimes is strongly inflowing or strongly outflowing. Figure~\ref{fig:sb_flows_z0} shows that the inflowing gas contributes significantly more strongly to the overall SB profile than the outflowing gas, and the inflows and slow-flow gas contribute roughly equally beyond 10 kpc, with slow-flow gas dominating within 10 kpc. This is not purely an effect of the amount of gas mass that fall into each cut: while the inflow selection does tend to contain $\sim2$--$6\times$ more mass than outflows across the six halos, the slow-flow selection contains $10$--$100\times$ more mass than the inflows despite contributing roughly equally to the \OVI\ SB profile. This means that observations of \OVI\ emission will likely probe either accreting material or slowly circulating material more than the fast outflows that are too hot for significant \OVI\ emission (see Section~\ref{subsec:foggie}).

Taken together with Figure~\ref{fig:sb_axes_z0}, which shows that there is no preferential direction for \OVI\ emission (i.e., along the major or minor axes), Figure~\ref{fig:sb_flows_z0} suggests that the inflows that contribute to the \OVI\ emission are \emph{not} primarily along the major axis. This suggests that the inflowing gas that is readily apparent as clumpy \OVI\ emission is not probing filamentary inflows that connect smoothly onto the galactic disk \citep[e.g.,][]{Dekel2006}. Instead, perhaps the clumpy \OVI\ emission is probing a phenomenon more similar to galactic fountains, where enriched material cycles from outflows back to inflows onto the galaxy, potentially after mixing with some fresh gas in the CGM \citep[e.g.,][]{Fraternali2013}. In particular, \citet{Augustin2025} found \OVI\ is likely enhanced in shells surrounding the densest clouds, which are colder in their cores and may be surrounded by a radiative mixing layer of intermediate-temperature gas \citep{Kwak2010,Stern2016,Ji2019,Liang2020}. A close examination of Maelstrom's \OVI\ emission map in Fig.~\ref{fig:maps_z0} reveals ``bubble"-like structures in the \OVI\ emission that could fit with this clump-shell picture. We examine the metallicity and enrichment of \OVI -traced gas in more detail in Section~\ref{subsec:feedback}, and further discuss how \OVI\ may be tracing a galactic fountain in Section~\ref{subsec:baryon_cycle}.

\OVI\ ions can be collisionally ionized or photoionized, with each ionization mechanism typically dominating at different temperature ranges. \OVI\ collisional ionization tends to peak in the range of $T\sim10^{5-6}$ K, while photoionization dominates at cooler temperatures around $T\sim10^4$ K \citep{Gnat2007,Oppenheimer2013a,Taira2025}. Figure~\ref{fig:phase} shows a temperature-density phase plot of all gas in the Maelstrom halo at $z=0$, color-coded by a histogram of \OVI\ emissivity. The emissivity peaks at temperatures between $T=10^5$ and $10^6$ K and densities $n\gtrsim10^{-4}$ cm$^{-3}$. There is some \OVI\ emissivity at lower temperatures ($T\sim10^4$ K), but the emissivity from gas at these temperatures is lower than at higher temperatures by a factor of $\sim2$ dex. The emission from these cooler phases tends to originate in more diffuse gas with number densities $\lesssim10^{-2}$ cm$^{-3}$, suggesting that cool \OVI\ emission is not located in dense clumps. Instead, cool \OVI\ emission could be originating in more diffuse structures---perhaps inflowing filamentary streams that are not much denser than the surrounding CGM \citep[C. Lochhaas et al., in prep.;][]{Strawn2021} or diffuse gas at large distances from the galaxies \citep{Stern2018}. The densest, coldest tail of the phase diagram is made up of gas located in and near the galaxy's interstellar medium, and there is no \OVI\ emission originating from this gas. We show just the Maelstrom halo at $z=0$ for simplicity, but the other halos have similar phase plots.

While our method of computing \OVI\ emissivity limits us to assuming \OVI\ is in photoionization and collisional ionization equilibrium, we can test the validity of this assumption and where it may break down. We evaluate ionization equilibrium using the ratio of ionization equilibration time to cooling time, $t_\mathrm{ion,eq}/t_\mathrm{cool}$. We follow \citet{Gnat2007} and define the timescale for \OVI\ to reach ionization equilibrium as
\begin{equation}
    t_\mathrm{ion,eq} = \frac{1}{\Gamma_{\rm{OVI}} + n_e \alpha_{\rm{OV}}}
\end{equation}
where $\Gamma_{\rm{OVI}}$ is the rate of ionization from \OVI\ to \ion{O}{7}, $\alpha_{\rm{OV}}$ is the rate of recombination to \ion{O}{5}, and $n_e$ is the electron density. We obtain the ionization and recombination rates from the same CLOUDY runs we use to make the emissivity maps. We take the cooling time as calculated from the simulation through the Grackle chemistry and cooling library, which follows non-equilibrium H and He chemistry but assumes solar relative abundances and equilibrium cooling for metals \citep[see Section~\ref{subsec:foggie};][]{Smith2017}. The term $t_\mathrm{ion,eq}$ is the timescale it takes for \OVI\ to return to ionization equilibrium if it is perturbed away from equilibrium. If this timescale is short compared to other relevant timescales, then we expect the assumption of \OVI\ ionization equilibrium to be satisfied. However, if this timescale is long compared to the timescale for perturbation, such as the cooling time $t_\mathrm{cool}$, then ionization equilibrium may be a poor assumption.

\begin{figure}
    \centering
    \includegraphics[width=\linewidth]{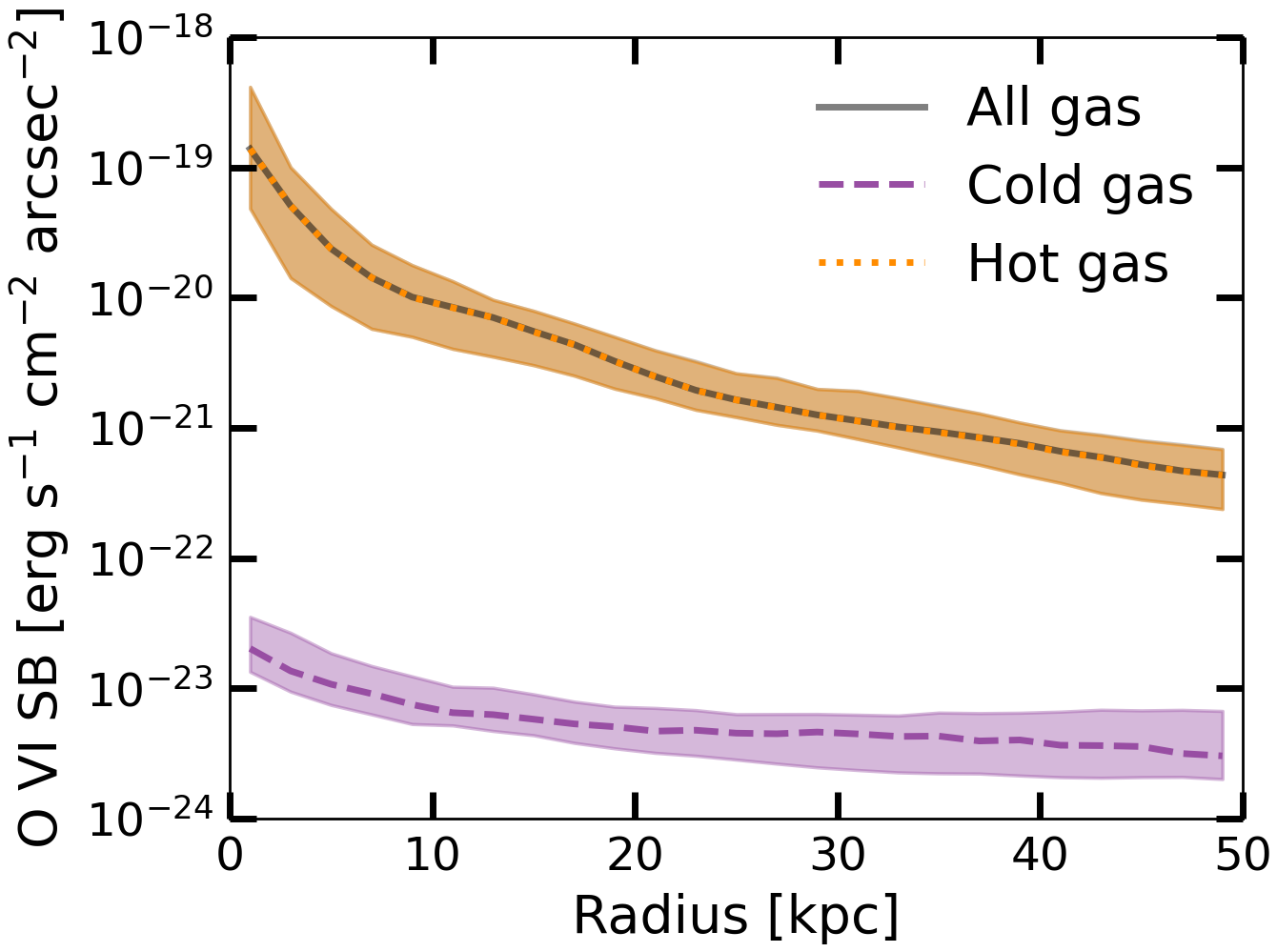}
    \caption{The average of the surface brightness profiles of all six galaxies' \OVI\ emission as a function of galactocentric distance (black curve) compared to the \OVI\ SB profile of only that gas that is hot with $T>10^5$ K (orange dotted curve) or cold with $T<10^5$ K (purple dashed curve), averaged over $z=0$--0.1. The black curve is the same as in Figure~\ref{fig:sb_axes_z0}. Shaded regions show the standard deviation of the SB profiles among the six simulated galaxies and over time. The vast majority of the \OVI\ surface brightness originates from hot gas.}
    \label{fig:sb_temp_z0}
\end{figure}

\begin{figure*}
    \centering
    \includegraphics[width=\linewidth]{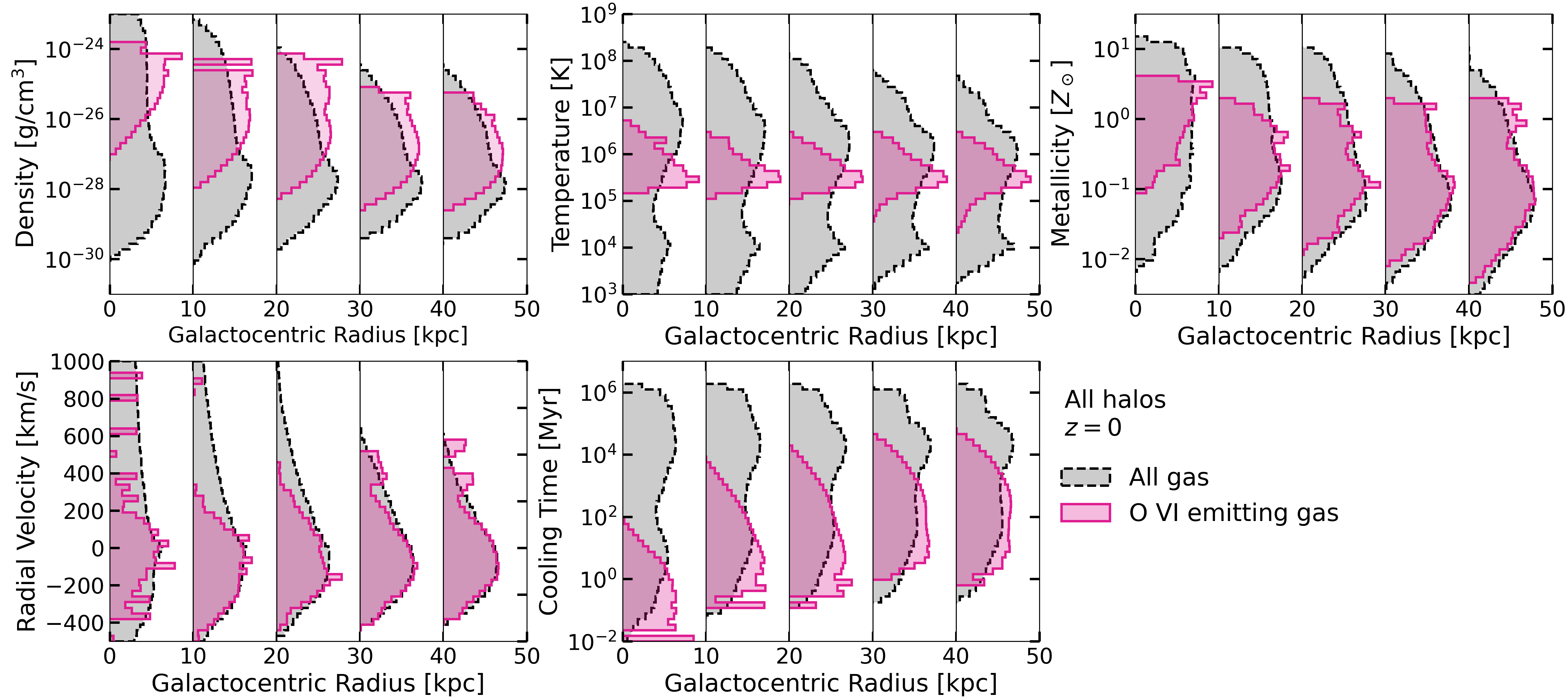}
    \caption{Histograms of gas properties in galactocentric radius bins, for all gas weighted by volume (black histograms) and weighted by \OVI\ volumetric emissivity (pink histograms), for all six halos at $z=0$. \OVI\ emission picks out gas that is generally skewed toward the denser end of the distribution, with temperatures $\sim10^5$--$10^6$ K, similar or metallicities as the full distribution, radial velocities skewed toward no-flow or inflow, and shorter cooling times than the full distribution.} \label{fig:gas-hists}
\end{figure*}

Figure~\ref{fig:tcool_teq} shows a projected image of $t_\mathrm{ion,eq}/t_\mathrm{cool}$ for the Maelstrom halo at $z=0$ on the left, and a phase plot of the same system color-coded by $t_\mathrm{ion,eq}/t_\mathrm{cool}$ on the right. Places where $t_\mathrm{ion,eq}/t_\mathrm{cool}$ approach or exceed 1 are places where the cooling time is too fast for the ionization to re-equilibrate, and therefore \OVI\ may be out of ionization equilibrium. $t_\mathrm{ion,eq}/t_\mathrm{cool}\gtrsim1$ occurs on the edges of \OVI-bright clouds and the edges of the extended gas disk surrounding the galaxy. The phase plot indicates these regions are at temperatures around $T\sim10^{4.5}$--$10^5$ K with a wide range of densities, $n\sim10^{-5}$--$10^{-2}$ cm$^{-3}$. The cooling curve peaks in efficiency around these temperatures, so it is unsurprising that the cooling times here are short.

If gas cools faster than it can recombine to lower ionization states of oxygen, then \OVI\ can be ``frozen in", producing higher ionization fractions than expected from the density and temperature of the gas \citep{Gnat2007,Kumar2025}. A similar phenomenon can also occur in galactic winds, if the wind expands faster than it can recombine \citep{Gray2019a,Chen2025}, and in cosmic noon galaxies with local radiation \citep{Cadiou2025}. While non-equilibrium effects make \OVI\ emission observations more challenging to model, they can lead to boosts in the \OVI\ emission SB by prolonging the lifetime of \OVI\ ions. The ionization equilibrium assumptions we use here therefore produce a lower limit on the \OVI\ SB, and the real CGM may have brighter emission than we predict.

To see the temperature dependence of \OVI\ emission more clearly, Figure~\ref{fig:sb_temp_z0} shows the \OVI\ surface brightness profile for all gas (black solid curve), gas with $T>10^5$ K (orange dotted curve), and gas with $T<10^5$ K (purple dashed curve). The contribution of cool gas to the \OVI\ surface brightness is lower than that from the warmer gas by a factor of $\sim2$ dex and up to $\sim3$ dex within 10 kpc from the galaxy center, suggesting that photoionized gas is strongly subdominant to collisionally ionized gas in the \OVI\ emission maps of Fig.~\ref{fig:maps_z0}.

While surface brightness profiles are more directly comparable to observations, the benefit of a 3D simulation is that we can directly examine the distribution of properties of the gas that contributes to \OVI\ emission. Figure~\ref{fig:gas-hists} shows histograms of gas density, temperature, metallicity, radial velocity, and cooling time. Each panel shows histograms within a galactocentric radial bin of radial width 10 kpc. The black histograms show properties of all gas in the simulation, weighted by volume\footnote{A mass weighting produces very similar histograms, with the strongest difference being a more prominent peak in the high-density part of the bimodal density distribution within 20 kpc of the galaxy.}. The pink histograms again show properties of all gas, but now weighted by \OVI\ volumetric emissivity. We combine information from all six halos at $z=0$.

\OVI\ emission is skewed towards gas with higher densities than the full distribution, but shifts toward lower densities with increasing distance from the galaxy. The density distribution of all gas becomes narrower with increasing radius, primarily losing the high-density side, but the peak of the distribution is consistent with radius out to 50 kpc. Close to the galaxy, \OVI\ emission does not extend to the densest regions because this dense gas is cold, where \OVI\ emissivity drops (see Fig.~\ref{fig:phase}). Regardless of distance from the galaxy, \OVI\ emission always peaks in the temperature range $10^5-10^6$ K, the range at which collisional ionization is the expected ionization mechanism. However, at large distances from the galaxy, a tail of \OVI\ emission arising from cooler gas down to $T\sim10^{4.5}$ K starts to develop, perhaps arising in cool and diffuse inflowing streams.

The metallicity distribution of \OVI-emitting gas is very similar to the metallicity distribution of all gas at most radii, although interestingly does not extend to the highest metallicites in the distribution. The interstellar medium of the FOGGIE galaxies is highly metal-enriched \citep{Acharyya2025}, but is too cold and dense for much \OVI\ emission, potentially explaining metallicity distribution differences close to the galaxy. Far from the galaxy, the \OVI-traced metallicity distribution matches the distribution of all gas much more closely. \OVI\ emission primarily arises in gas with radial velocities $\lesssim200$ km s$^{-1}$, peaking at negative velocities (indicating inflow), but the \OVI-traced gas distribution develops a tail toward large positive velocities far from the galaxies, indicating that there is some (faint) \OVI\ emission located in the fast outflows. At smaller radii, the high-velocity (outflow) tail in the full distribution contains gas too hot for \OVI\ emission. Finally, we also see that \OVI\ emission picks out gas with the shortest cooling times, but the cooling time of \OVI-emitting gas increases with increasing galactocentric radius, tracking a similar increase in the cooling time distribution of all gas. Cooling time is inversely proportional to density so as the high-density side of the density distribution disappears at larger radii, it is expected that cooling time will increase accordingly. Because \OVI\ emission peaks at a temperature range generally considered to be a ``transition temperature", where radiative cooling is highly efficient, it is not surprising that it arises from gas with short cooling times.

\subsection{Trends with Halo Properties} \label{subsec:halo_prop}

\begin{figure*}
    \centering
    \includegraphics[width=0.85\linewidth]{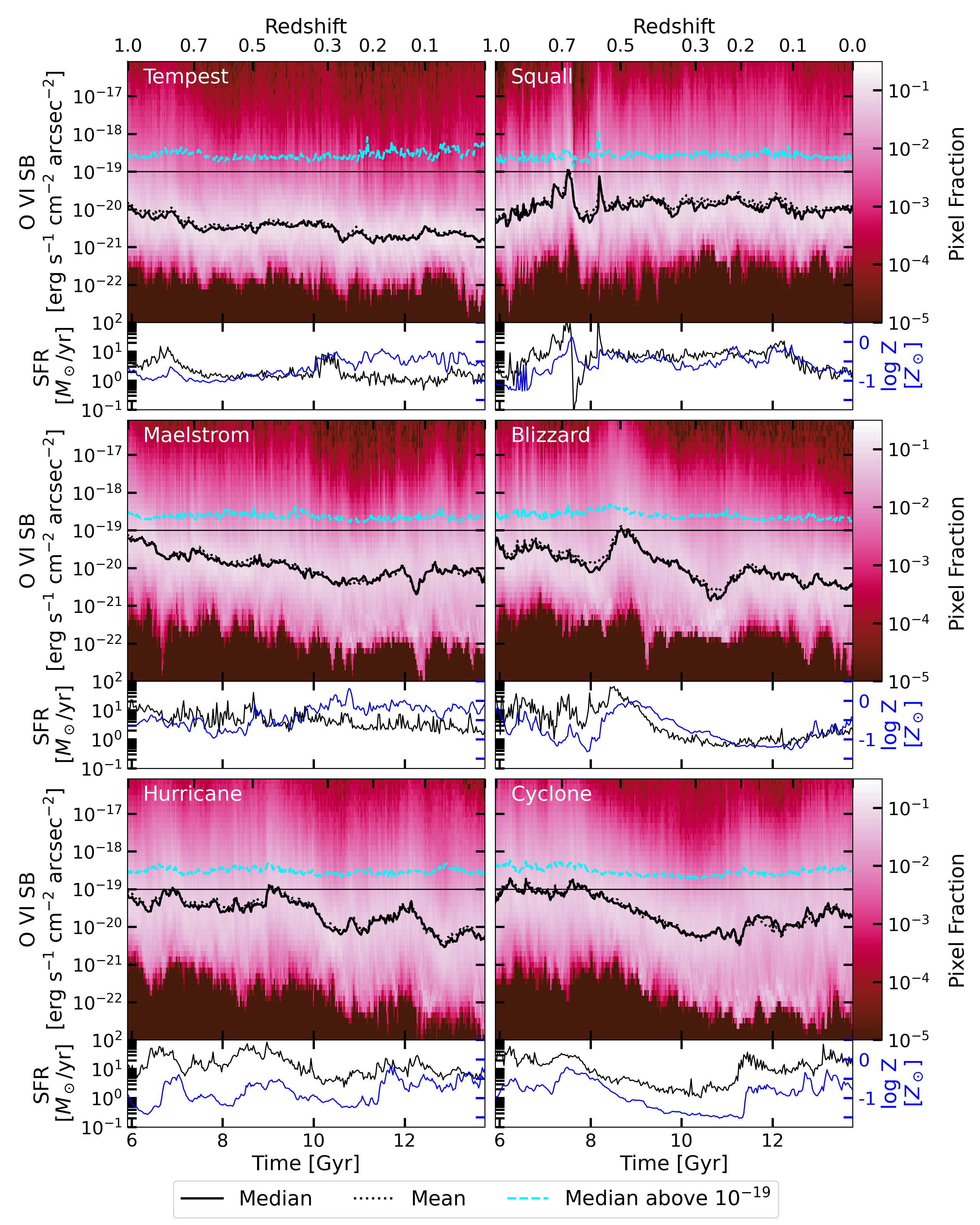}
    \caption{Top of each panel shows histograms of \OVI\ surface brightness over time from $z=1$ to $z=0$. The color coding in each vertical column shows a histogram of \OVI\ surface brightnesses for simulation cells within 20 kpc of the center of the galaxy, projected edge-on. Brighter colors at a certain surface brightness on the vertical axis indicate more cells with that surface brightness. The solid black curve shows the median value of \OVI\ surface brightness, the dotted black curve shows the mean, and the dashed cyan curve shows the median value of \OVI\ surface brightness when considering only those pixels with a surface brightness above a limit of $10^{-19}$ ergs s$^{-1}$ cm$^{-2}$ arcsec$^{-2}$. The \OVI\ emission is calculated assuming a $z=0$ extragalactic UV background even at higher redshifts to separate the effect of the background from halo and galaxy effects. There are significant variations in the typical surface brightness and the spread of brightnesses over time and a weak trend toward lower typical surface brightnesses at later times. Bottom of each panel shows the star formation rate of the galaxy (black, left axis) and the median CGM metallicity within 20 kpc of the galaxy (blue, right axis) at the same time snapshots as in the upper panels. Each panel shows a different FOGGIE galaxy.} \label{fig:sb_vs_time}
\end{figure*}

The FOGGIE simulation suite contains only six $\sim L^\star$ galaxies that were selected to be somewhat similar to each other in halo mass and relatively isolated at $z=0$. The vast differences in the \OVI\ emission morphology among these similar galaxies revealed by Fig.~\ref{fig:maps_z0} suggests that if there are any trends in \OVI\ morphology or brightness with galaxy properties, they may be slight and possibly obscured by large amounts of intrinsic variation. To better probe correlations between \OVI\ brightness and galaxy properties, we examine the properties of the galaxies and the predicted emission as the simulated galaxies evolve over cosmic time. We continue to use the $z=0$ UVB when calculating \OVI\ emission from the higher-$z$ outputs, such that the different snapshots in time for a given galaxy can be thought of as additional $z\sim0$ snapshots. We begin with investigating the \OVI\ surface brightness trends directly with redshift to determine if there are time evolution trends to consider (even when using the same $z=0$ UVB at all times), then move on to trends with halo mass, average CGM density, temperature, metallicity, and finally star formation rate (SFR) of the central galaxy.

Figure~\ref{fig:sb_vs_time} shows histograms of the fraction of pixels within 20 proper kpc of the center of the galaxy in edge-on projections (same viewing angle as in Fig.~\ref{fig:maps_z0}) in the 2D space of \OVI\ SB (vertical axis) and cosmic time (horizontal axis) from $z=1$ to $z=0$, with time spacing of $\sim26$ Myr between snapshots (293 total snapshots for each galaxy). We choose the region within 20 kpc of the center of the galaxy to focus on because this is the maximum extent of likely detectable \OVI\ emission (see Section~\ref{subsec:Aspera}) for any of the FOGGIE galaxies at $z=0$. Each panel shows the evolution of a different FOGGIE galaxy, with black solid (dotted) curves indicating the median (mean) surface brightness within 20 kpc as a function of time. There is significant variation over time and between galaxies, but four of the six galaxies show a noisy and shallowly decreasing trend of median \OVI\ SB by about 0.5--1 dex over the time range shown. The bottom panels underneath each 2D histogram show the star formation rate (SFR) of the central galaxy in black (left axis) and the median metallicity of CGM gas within 20 kpc of the galaxy center in blue (right axis) over the same span of cosmic time. The SFRs of the FOGGIE galaxies are generally burstier at high redshift, and for many of the strong starbursts, a large fraction of the pixels shift to a larger \OVI\ SB (upwards in Fig.~\ref{fig:sb_vs_time}) and both the median \OVI\ SB and the median CGM metallicity increases around, or shortly after, the time of the burst. Cyclone is an exception, as it has a much higher star formation rate near $z=0$ than the other galaxies, and correspondingly does not show the same decrease in median \OVI\ SB over time as the other galaxies. This is a first clue that \OVI\ SB is correlated with SFR and somewhat also CGM metallicity, which will be explored more below.

There are a few examples where the \OVI\ SB time evolution does not appear to be linked to the star formation history. Blizzard exhibits a drastic decline in the \OVI\ SB at $\sim11$ Gyr ($z\sim0.25$), when the SFR is fairly constant at a low $\sim1M_\odot$ yr$^{-1}$. This galaxy experienced a gas-rich merger at 8.75 Gyr ($z=0.55$) that boosted the SFR, which then launched strong outflows that nearly quenched the galaxy for several billion years. During this time, when the SFR is low, the CGM metallicity dips (see Fig.~\ref{fig:prop_vs_t} in Appendix~\ref{app:time_depend}). It is this low CGM metallicity that is likely contributing to the decline in \OVI\ SB until the galaxy begins to rejuvenate and start populating the CGM with metals again. Maelstrom and Tempest both also show smaller, similar events unassociated with the SFR, where the \OVI\ SB suddenly dips for a short length of time before partially or fully recovering, at $\sim12.5$ Gyr ($z\sim0.1$) for Maelstrom and at $\sim11$ Gyr ($z\sim0.2$) for Tempest. In these cases, the median CGM metallicity is still high (Fig.~\ref{fig:prop_vs_t} in Appendix~\ref{app:time_depend}) but the median CGM gas density has dropped. Because \OVI\ emission depends on both metallicity and gas density, as we saw in Section~\ref{subsec:emiss_props}, lower values of these gas properties are responsible for lower \OVI\ SB values. The SFR can impact median CGM densities and metallicities, as can accretion and orbiting satellites, so it is unsurprising that this translates into a link between \OVI\ SB and SFR, albeit imperfect.

Upcoming CGM emission instruments will detect emission only above their sensitivity threshold. Here, we approximate a sensitivity threshold for \OVI\ as $10^{-19}$ ergs s$^{-1}$ cm$^{-2}$ arcsec$^{-2}$, similar to that expected for deep observations by the upcoming \emph{Aspera} mission \citep{Chung2021b}. The horizontal black lines in Fig.~\ref{fig:sb_vs_time} indicate this threshold, and the dashed cyan line shows the median \OVI\ SB when only pixels above this limit are detected. Interestingly, the median SB of detected pixels does not show the 0.5--1 dex decrease over redshift that the median SB of all pixels shows, nor is it as variable over time or as correlated with SFR. This suggests that the high-SB end (i.e., the observable end) of the emission maps does not change significantly over redshift.

Near-future CGM UV emission instruments are unlikely to observe \OVI\ emission in galaxies with redshifts higher than $z\sim0.01$, but the small amount of variation over time, and similarity between the six halos in the median SB, allows us to treat the different time snapshots of each galaxy as representative of $z\approx0$ snapshots. The FOGGIE suite contains just six galaxies, so including additional snapshots at different times allows us to fill in any correlations with more data, as long as we understand how the properties we explore depend on time (see Appendix~\ref{app:time_depend}).

\begin{figure*}
    \centering
    \includegraphics[width=0.49\linewidth]{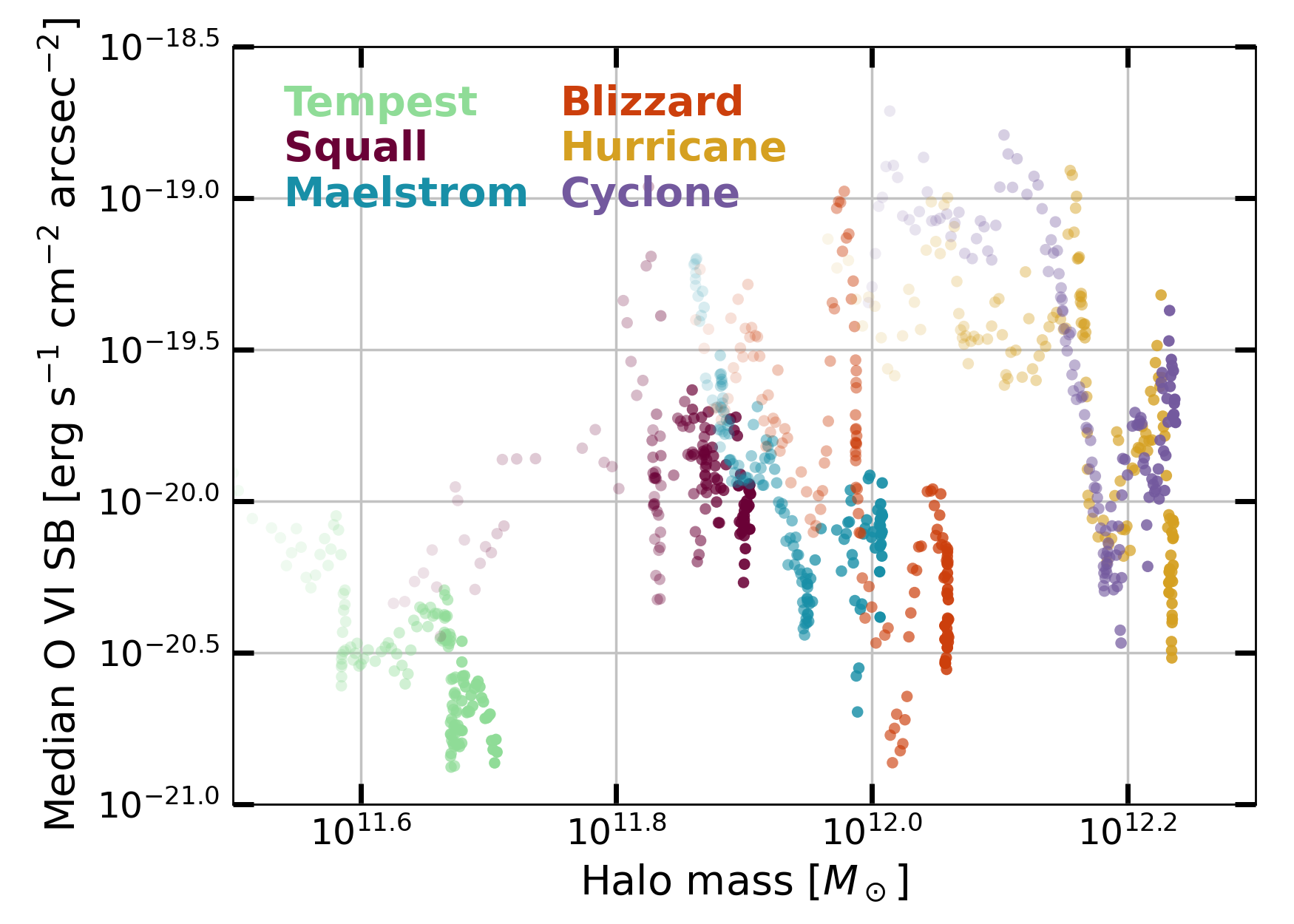}
    \includegraphics[width=0.49\linewidth]{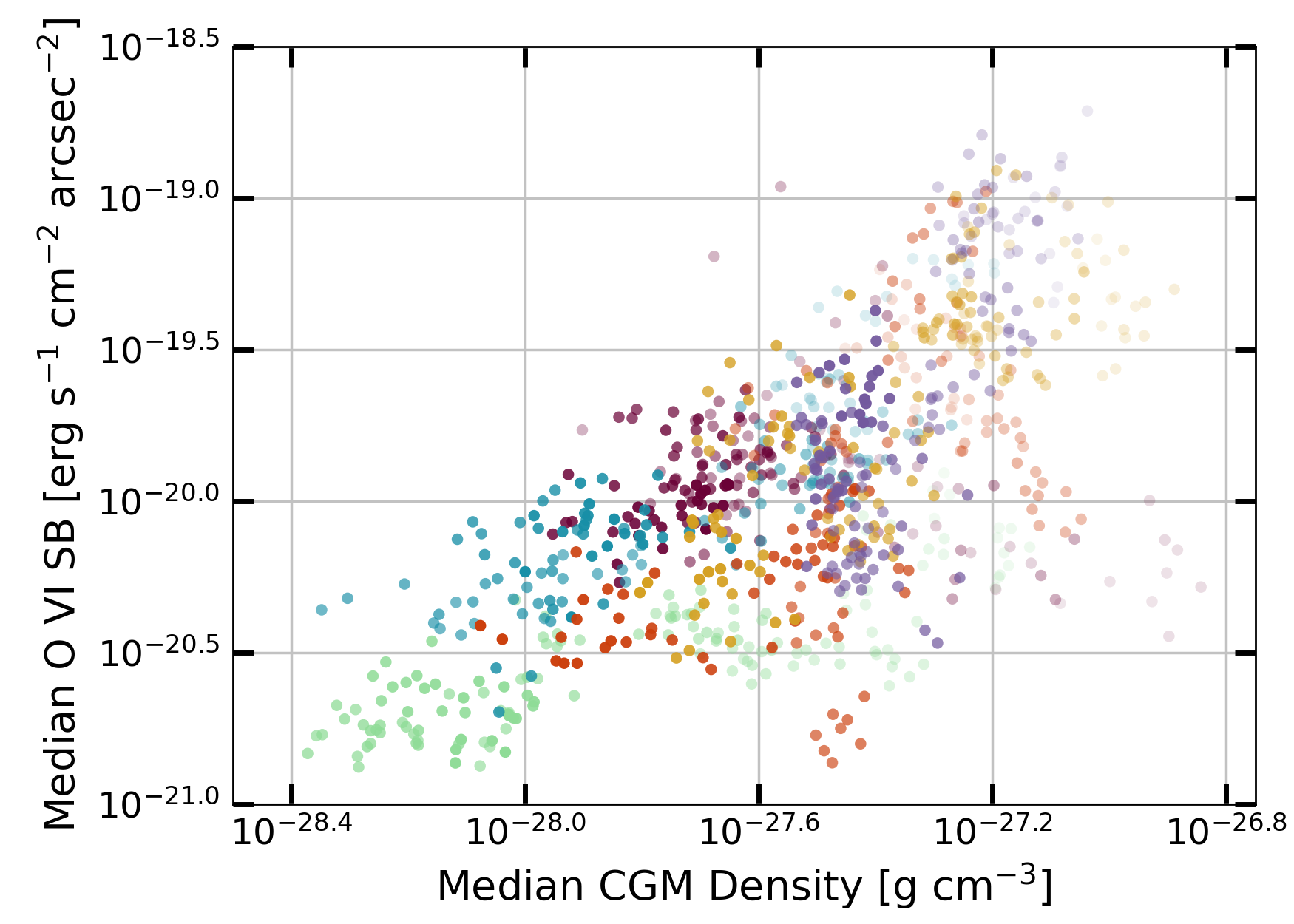}
    \includegraphics[width=0.49\linewidth]{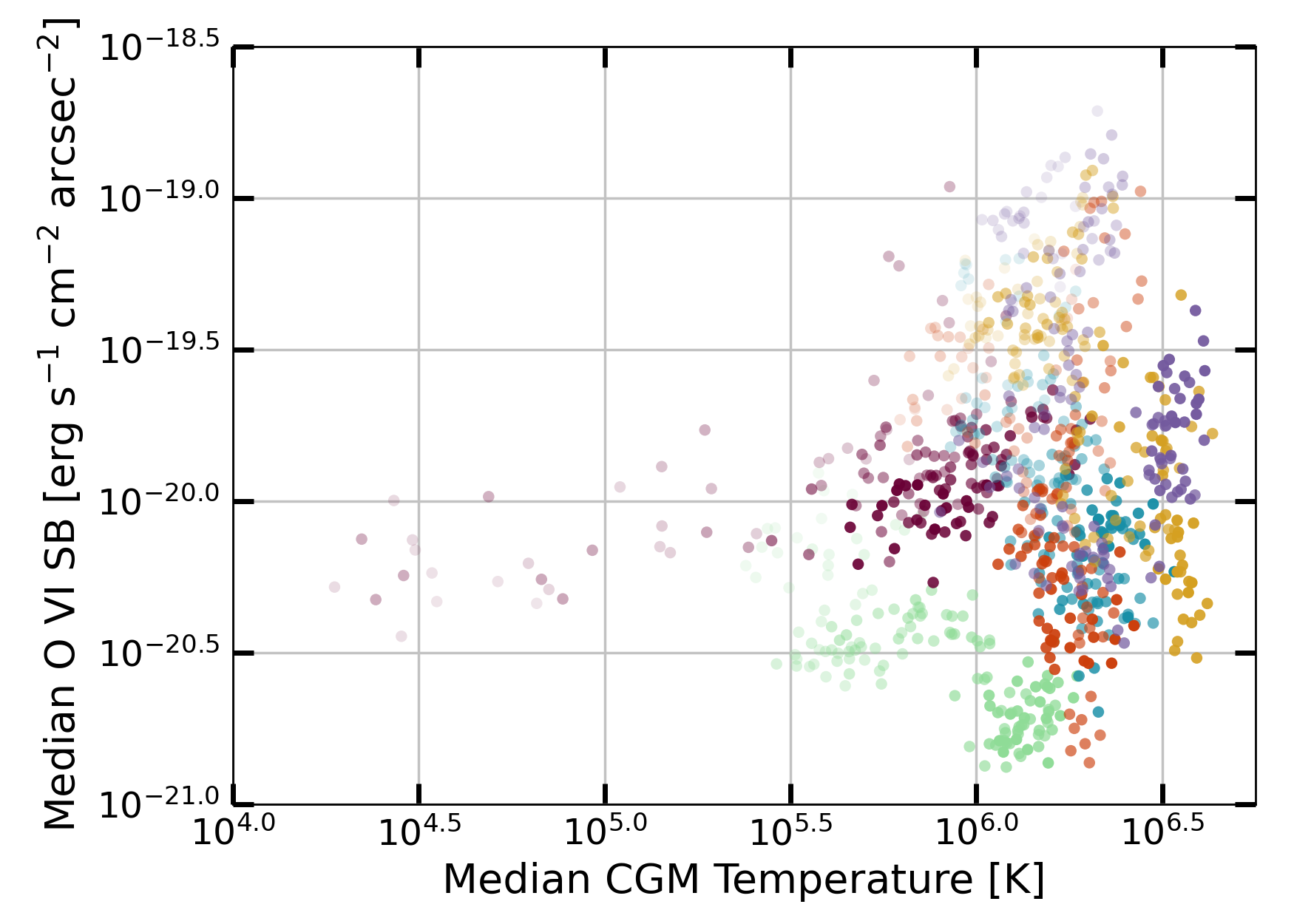}
    \includegraphics[width=0.49\linewidth]{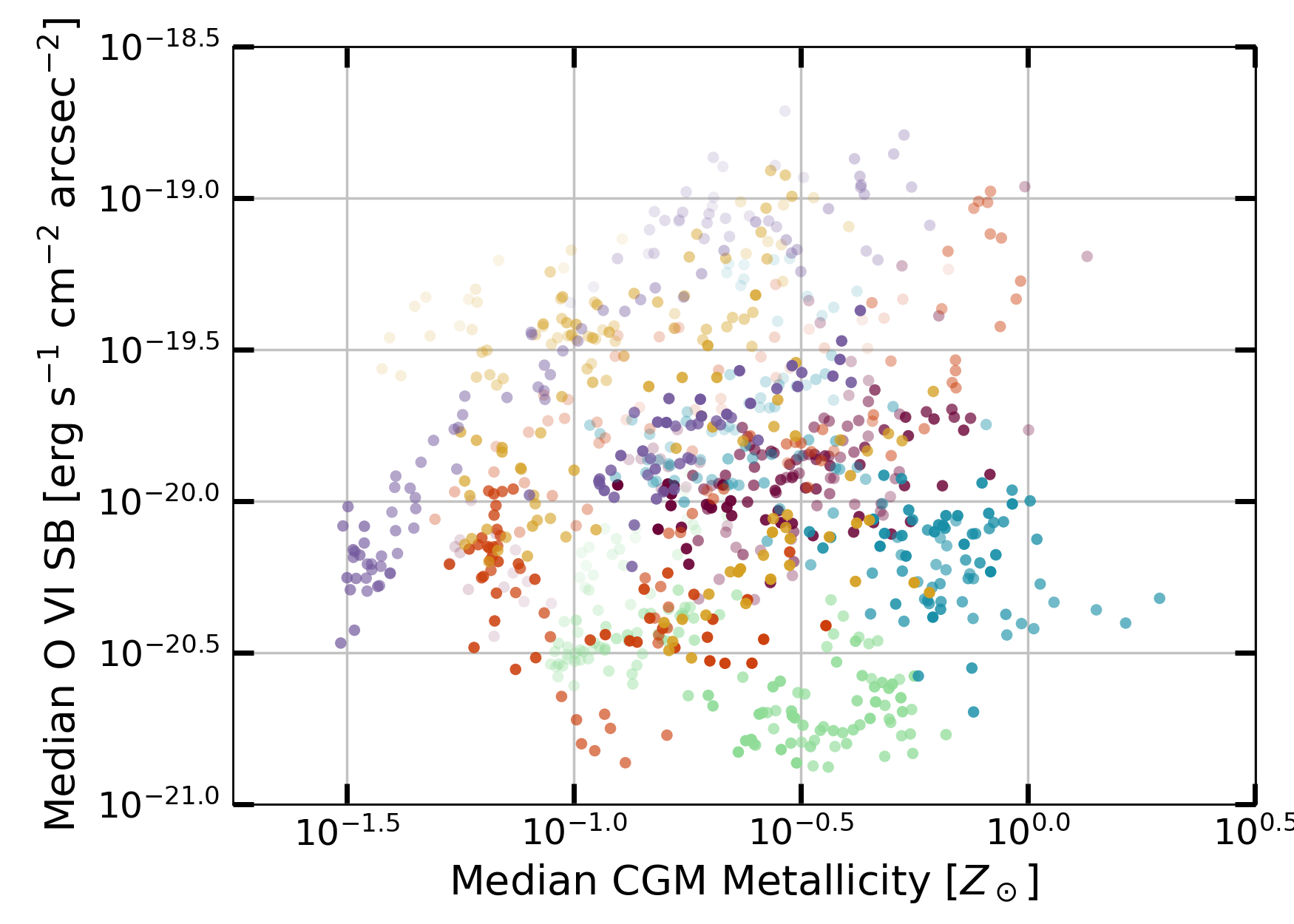}
    \caption{The median \OVI\ SB within 20 kpc of the galaxy center, projected edge-on, as a function of the total mass (stars, gas, and dark matter) of each galaxy's halo within its virial radius (top left), median CGM density (top right), temperature (bottom left), or metallicity (bottom right) within the same 20 kpc from the galaxy center. Marker opacity indicates the snapshot in time between $z=1$ (fainter) and $z=0$ (bolder), and the marker color indicates the FOGGIE galaxy as in the legend. The strongest trend is with average CGM density, followed by halo mass, and no noticeable trends with average CGM temperature or metallicity.}
    \label{fig:sb_vs_Mh-den-temp-Z}
\end{figure*}

\begin{figure}
    \centering
    \includegraphics[width=\linewidth]{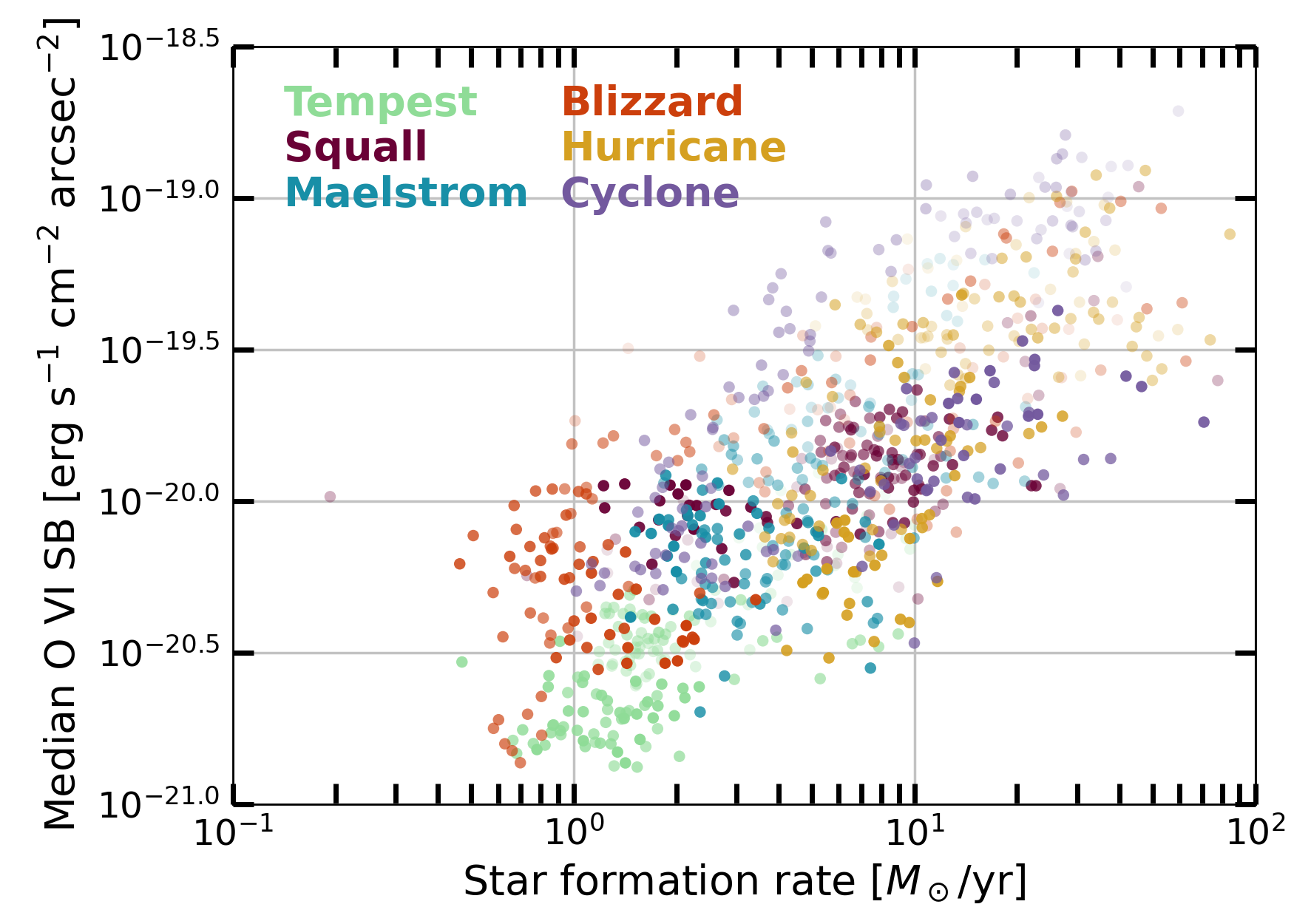}
    \caption{The median \OVI\ SB within 20 kpc of the galaxy center, projected edge-on, as a function of the SFR of the central galaxy. Point opacity represents time snapshots, between $z=1$ (faint) and $z=0$ (bold). Point color indicates each galaxy as in the legend. There is a clear trend of increasing median \OVI\ SB with increasing SFR.}
    \label{fig:sb_vs_sfr}
\end{figure}

To investigate what CGM or galaxy properties drive the general trend of decreasing \OVI\ SB with time shown by Fig.~\ref{fig:sb_vs_time}, we compare the median \OVI\ SB to various CGM and galaxy properties. We again use the median \OVI\ SB within 20 kpc of the galaxy center, projected edge-on as in Fig.~\ref{fig:maps_z0}. Figure~\ref{fig:sb_vs_Mh-den-temp-Z} shows the median \OVI\ SB as functions of halo mass and median CGM density, metallicity, and temperature within the same 20 kpc galactocentric radius. While the \OVI\ SB is derived from the projected images (in the same edge-on orientation as Fig.~\ref{fig:maps_z0}), we define the median CGM properties as the volume-weighted median of gas cells within a 20 kpc spherical region centered on the galaxy. To ensure we capture primarily CGM rather than interstellar medium (ISM) properties, we use the time-evolving gas density cut of \citet{Lochhaas2023} to remove each galaxy's gas disk before computing the median of the gas properties\footnote{This density cut does remove some of the denser clumps of gas in the CGM, but because the medians are volume-weighted and the dense clumps are small, this does not strongly affect the median values.}, but continue to calculate the median \OVI\ SB from all gas. Each point in these figures shows the \OVI\ SB and galaxy or CGM properties at a single snapshot in time, where fainter points indicate earlier times closer to $z=1$ and bolder points indicate later times closer to $z=0$. We use 146 snapshots, every $\sim53$ Myr between $z=1$ and $z=0$, for each of the six FOGGIE galaxies, where each galaxy is assigned a different marker color. We decrease the time cadence for Figures~\ref{fig:sb_vs_Mh-den-temp-Z}-\ref{fig:sb_vs_sfr} to snapshots every $\sim53$ Myr instead of every $\sim26$ Myr as in Fig.~\ref{fig:sb_vs_time} to reduce clutter. The galaxies' halo masses grow over time, and the average CGM density decreases over time, but the average CGM temperature and metallicity are more time-variable with only a very weak general increase over the time range from $z = 1$ to $z = 0$ (see Appendix~\ref{app:time_depend}).

The top left panel of Figure~\ref{fig:sb_vs_Mh-den-temp-Z} shows the median \OVI\ SB as a function of total halo mass, including gas, stars, and dark matter within the virial radius for each galaxy. The four galaxies that exhibited a down-turn in the \OVI\ median brightness with cosmic time in Fig.~\ref{fig:sb_vs_time} (all but Squall and Cyclone) also exhibit this down-turn at their individual largest halo masses, as the largest halo mass generally corresponds with the latest simulated time ($z=0$). Other than the lowest-mass galaxy (Tempest) generally having lower \OVI\ median SBs than the rest of the galaxies, there does not appear to be a significant trend of median \OVI\ SB with increasing halo mass. The FOGGIE galaxies do not span a large range in halo mass, so it is unsurprising that any halo mass trends are weak.

The top right panel of Figure~\ref{fig:sb_vs_Mh-den-temp-Z} shows the median \OVI\ SB as a function of median CGM gas density for all six FOGGIE galaxies. There is a trend of increasing \OVI\ SB with increasing median CGM density. Because emission scales with density squared, this trend is not surprising, but the slope of this relation is significantly shallower than the expected scaling with density squared---instead, it appears to be roughly proportional to the density. The shallow slope is likely because the plotted relation is between the median \OVI\ SB within 20 kpc of the galaxy center and the median CGM density in the same region, a more global relation than the direct density squared scaling of emissivity. The CGM density is generally larger at higher redshifts, and decreases toward low redshift, so the trend may explain the decrease in \OVI\ surface brightness over time seen in Fig.~\ref{fig:sb_vs_time}.

The bottom panels of Figure~\ref{fig:sb_vs_Mh-den-temp-Z} show the median \OVI\ SB as a function of median CGM temperature (left) and metallicity (right). At first glance, there does not appear to be a strong trend of \OVI\ emission with either median CGM metallicity or temperature. \OVI\ emissivity is expected to scale linearly with metallicity, as increasing metallicity increases the number of oxygen atoms that may be ionized to \OVI. However, much like the relation with median CGM density (top right panel), this scaling is expected only on a cell-by-cell basis and does not appear to translate to median values in the large 20 kpc radius sphere centered on the galaxy. The ion fraction for \OVI\ peaks at a temperature around $10^{5.5}$ K, so we might expect to see an increase in \OVI\ SB around that temperature and decreasing \OVI\ at both smaller and larger temperatures, but there is no clear relationship between median \OVI\ and median CGM temperature. This is likely because the median CGM temperatures within 20 kpc of the galaxy center are between $\sim10^{5.5}$--$10^{6.5}$ K for the FOGGIE galaxies, where \OVI\ ionization fractions in collisional ionization peak (see Section~\ref{subsec:emiss_props}), so \OVI\ emission picks out the peak of the temperature distribution for the CGM gas.

Figure~\ref{fig:sb_vs_sfr} shows the median \OVI\ SB vs. the SFR of the central galaxy. The SFR is defined as the mass of young stars with ages less than $10$ Myr within 20 kpc of the galaxy center divided by $10$ Myr. As in Figure~\ref{fig:sb_vs_Mh-den-temp-Z}, snapshots with $0<z<1$ are plotted for each of the six FOGGIE galaxies, with point color indicating each galaxy. There is a clear trend of increasing \OVI\ SB with increasing SFR: more strongly star-forming galaxies have brighter \OVI\ emission in their inner CGM. This is consistent with the features seen in Fig.~\ref{fig:sb_vs_time} where some peaks in the SFR appeared to be correlated with shifts in the \OVI\ SB distribution to larger values.

The \OVI\ SB exhibits an obvious correlation with SFR, but not with CGM metallicity (bottom right panel of Fig.~\ref{fig:sb_vs_Mh-den-temp-Z}), even though the bottom panels of Fig.~\ref{fig:sb_vs_time} show some correlations in time between SFR and metallicity. The time correlation between SFR and metallicity is not perfect---strong bursts or dips in the SFR correlate with increases and decreases in the median CGM metallicity, but relatively constant star formation histories, like that of Tempest and Maelstrom, do not appear to be correlated with the CGM metallicity. The lack of correlation for these two halos can also be seen in the bottom right panel of Fig.~\ref{fig:sb_vs_Mh-den-temp-Z}, where the mint green (Tempest) and blue (Maelstrom) clouds of points do not exhibit a correlation between median \OVI\ SB and CGM metallicity, but the other halos' clouds of points appear to present a weak correlation with large scatter. This suggests that inner CGM enrichment also plays a role in boosting \OVI\ brightness, but perhaps not as directly as the SFR. The apparent correlation (or lack thereof, for Tempest and Maelstrom) between SFR and metallicity is highly model-dependent, as it depends on how metals are deposited and dispersed by the feedback scheme.

A correlation with SFR may suggest that the stellar feedback, in the form of hot outflows launched from the galaxy, is responsible for boosting the \OVI\ SB. However, taken with Fig.~\ref{fig:sb_flows_z0} that shows it is primarily the inflows and slow-flow gas, and not the outflows, that contribute to the \OVI\ SB profile, it is peculiar that increases in the SFR drive increases in \OVI\ SB. Instead, it could be that the strong bursts of stellar feedback that increase the inner CGM metallicity are also enriching the inflowing gas that then glows in \OVI\ emission as it falls back to the galaxy. This suggests that \OVI\ emission could be a direct tracer of galactic fountain flows. We further discuss the effects of stellar feedback in the next section, and the ability of \OVI\ to trace fountains in Section~\ref{subsec:baryon_cycle}.

\subsection{Impact of Star Formation Feedback} \label{subsec:feedback}

There is a clear correlation between SFR and \OVI\ SB, but it is not yet clear what role the stellar feedback plays in this correlation\footnote{A similar correlation is seen in \OVI\ absorption studies, see Section~\ref{subsec:baryon_cycle}.}. In FOGGIE, the stellar feedback is in the form of purely thermal energy injection, where a fraction of $10^{-5}$ of the rest mass energy of new stars is returned to the gas as thermal energy (this roughly corresponds to $10^{51}$ ergs per 100 M$_{\odot}$ of stars formed). This feedback mechanism generates hot and fast galactic outflows with $T\sim10^{6-7}$ K and $v\sim1000$--2000 km s$^{-1}$. These temperatures are significantly higher than the temperature range where \OVI\ ion fractions peak, and such hot gas is also generally quite diffuse, so it is not surprising that the outflows do not contribute significantly to the \OVI\ SB profiles (Fig.~\ref{fig:sb_flows_z0}).

\begin{figure*}
    \centering
    \includegraphics[width=0.4\linewidth]{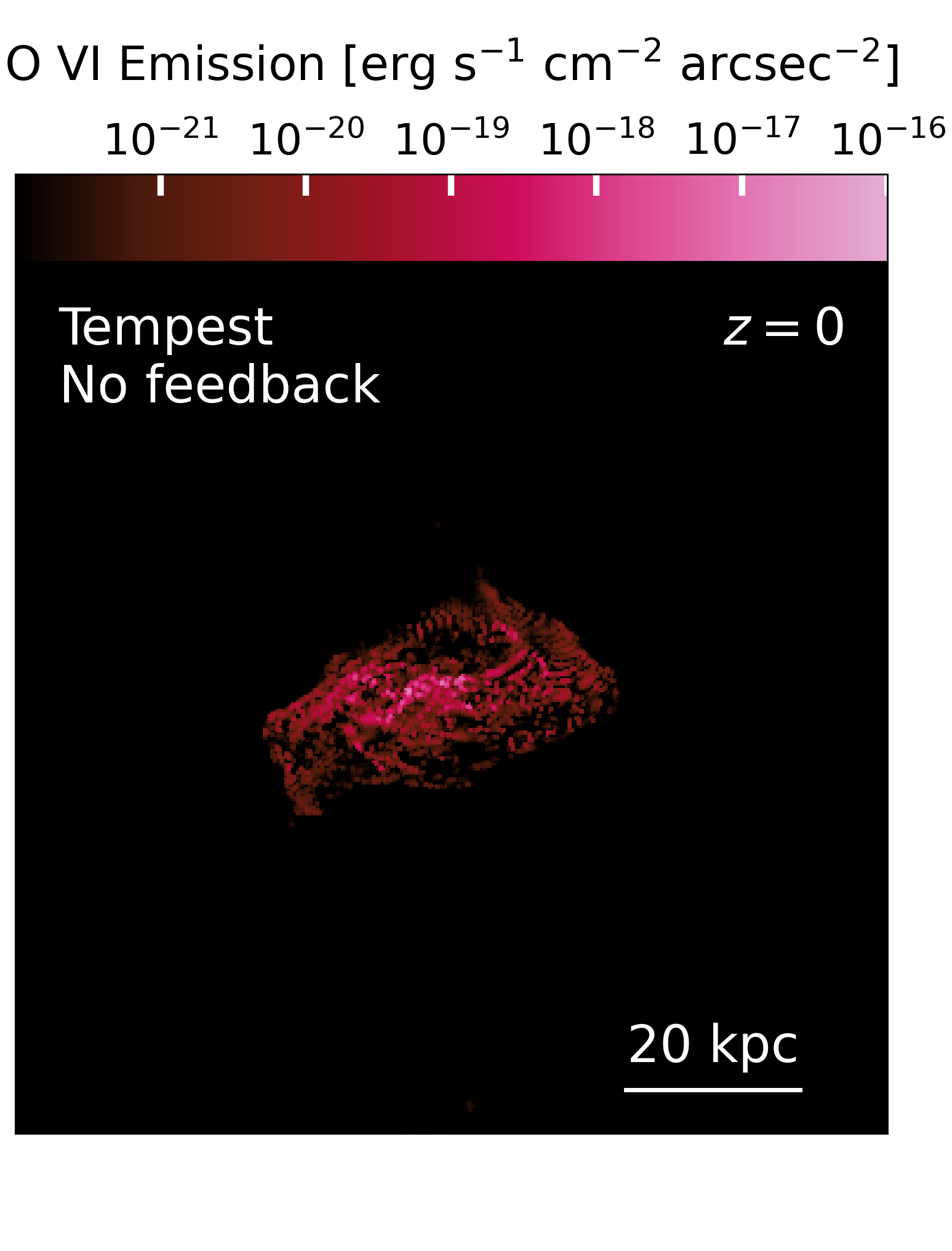}
    \includegraphics[width=0.59\linewidth]{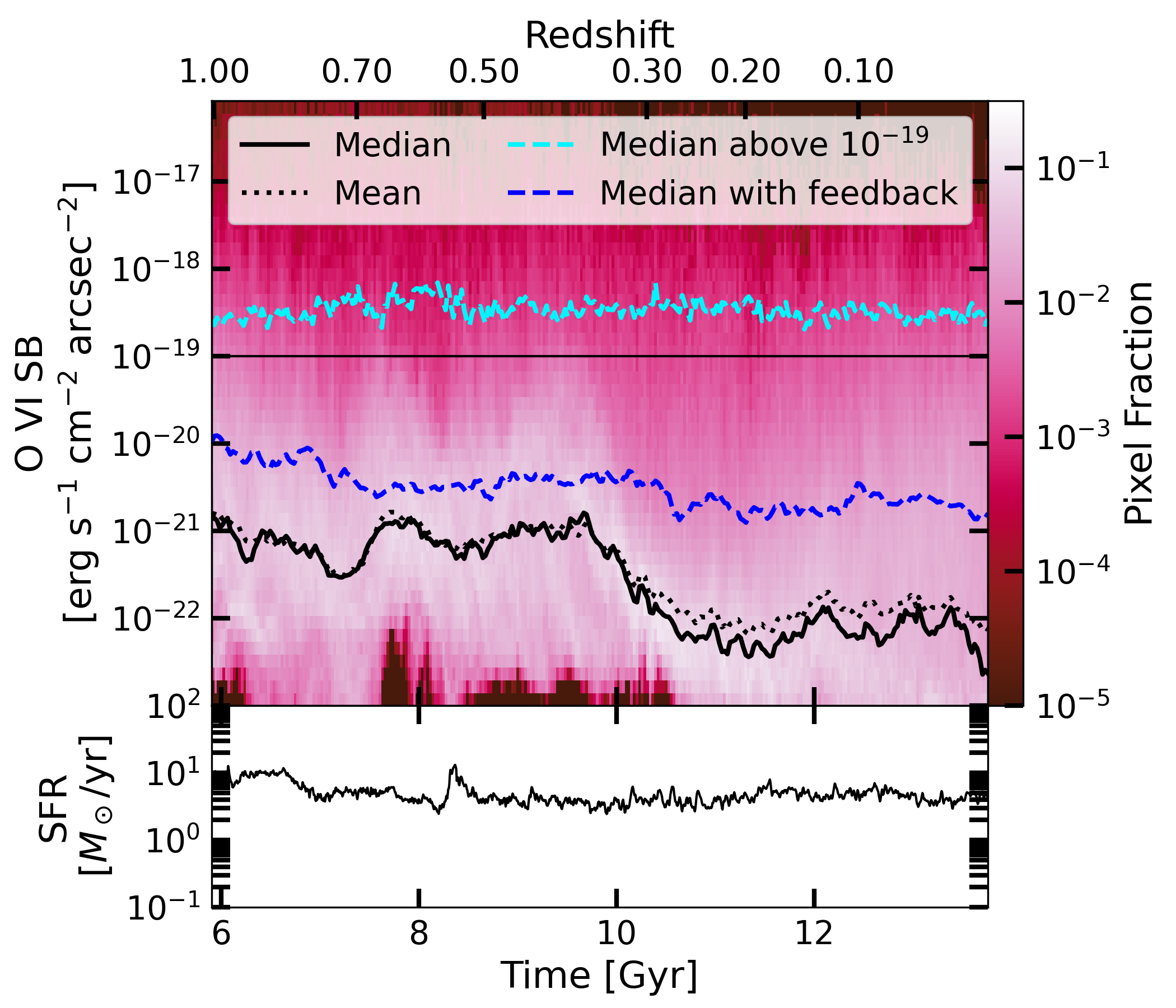}
    \caption{Left: An \OVI\ emission map of the no-feedback run of Tempest. Without feedback, the \OVI\ emission in the CGM is significantly reduced as compared to fiducial Tempest in Fig.~\ref{fig:maps_z0}. Right: A 2D histogram of pixel fraction in \OVI\ SB and time for the no-feedback run of Tempest (top panel) and the SFR over the same time (bottom panel), which can be compared to fiducial Tempest in Fig.~\ref{fig:sb_vs_time}. The black solid (dotted) curve shows the median (mean) \OVI\ SB within 20 kpc of the center of the galaxy at each time, which is $1-1.5$ dex lower than the fiducial run of Tempest (blue dashed curve). The cyan dashed curve shows the median when considering only pixels above a detection limit of $10^{-19}$ ergs s$^{-1}$ cm$^{-2}$ arcsec$^{-2}$.} \label{fig:no-feedback}
\end{figure*}

If fresh inflowing gas is what dominates the \OVI\ emission, then the strength of the stellar feedback should not affect the \OVI\ SB profiles. To test this hypothesis, we restarted the Tempest FOGGIE galaxy from the simulation initial conditions with the thermal energy injection set to $10^{-10}$ of the rest mass energy of new stars formed, essentially shutting off feedback entirely (although stars still returned metals to their immediate surroundings). Without feedback, simulated galaxies form too many stars \citep{Somerville2015}, and the no-feedback rerun of Tempest is no exception---it forms $1.7\times$ more stellar mass by $z=0$ than the fiducial Tempest. Despite the fact that Tempest without feedback does not form a ``realistic" galaxy, we can still investigate the impact on \OVI\ emission in the inner CGM to contrast with the fiducial run of Tempest and infer the effects of the stellar feedback.

Figure~\ref{fig:no-feedback} shows an \OVI\ emission map for the no-feedback run of Tempest (left) and the 2D histogram of pixel fraction in \OVI\ SB and time (right) in the same style as Fig.~\ref{fig:sb_vs_time}. It is immediately obvious that turning off the stellar feedback results in greatly reduced \OVI\ SB. The median \OVI\ (black curve in right panel) is $\sim2$ dex lower by $z=0$ than in the fiducial run of Tempest (Fig.~\ref{fig:sb_vs_time}), and visually the \OVI\ emission is weaker and more confined to the very inner region close to the galaxy (left panel).

\begin{figure*}
    \centering
    \includegraphics[width=0.49\linewidth]{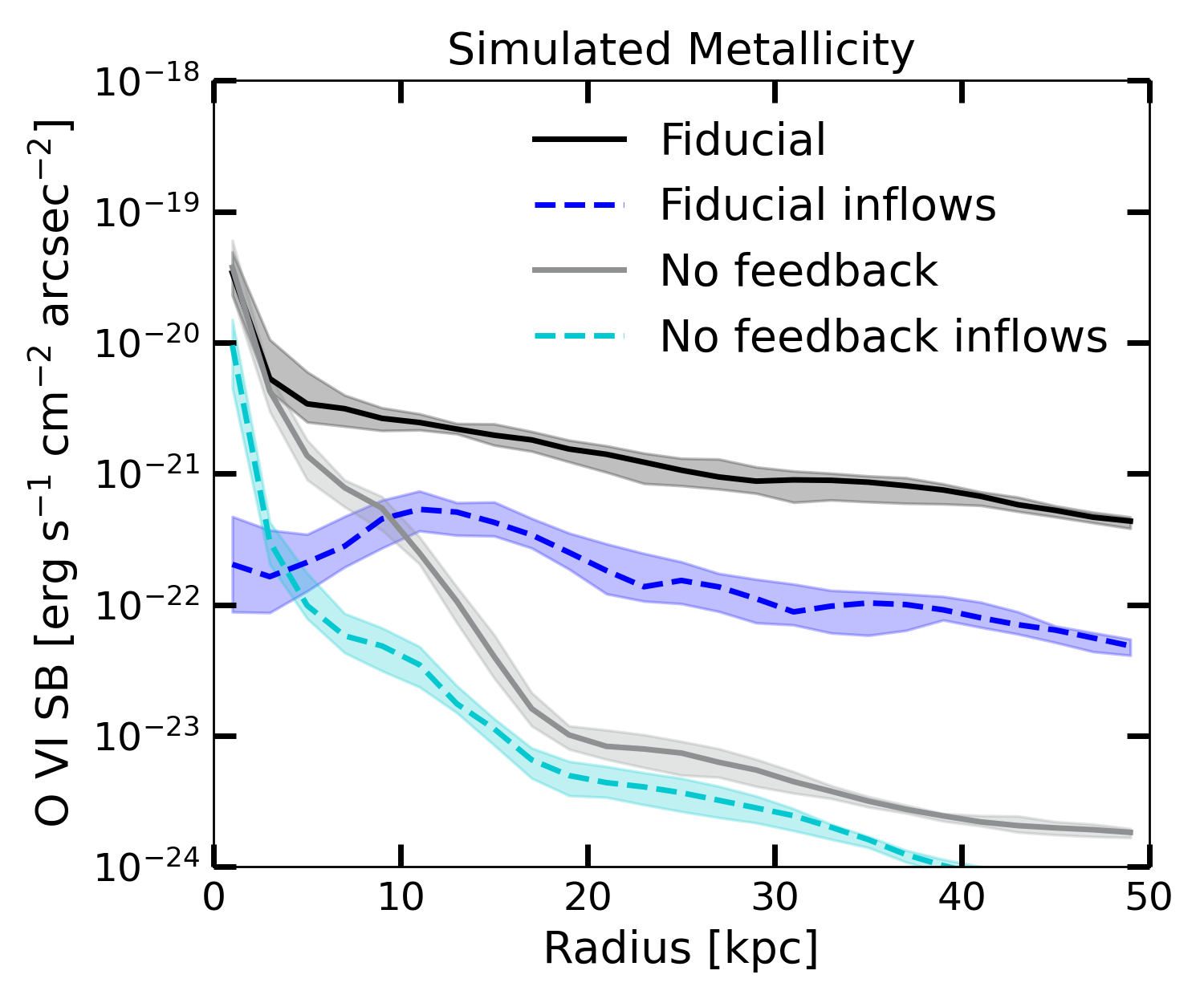}
    \includegraphics[width=0.49\linewidth]{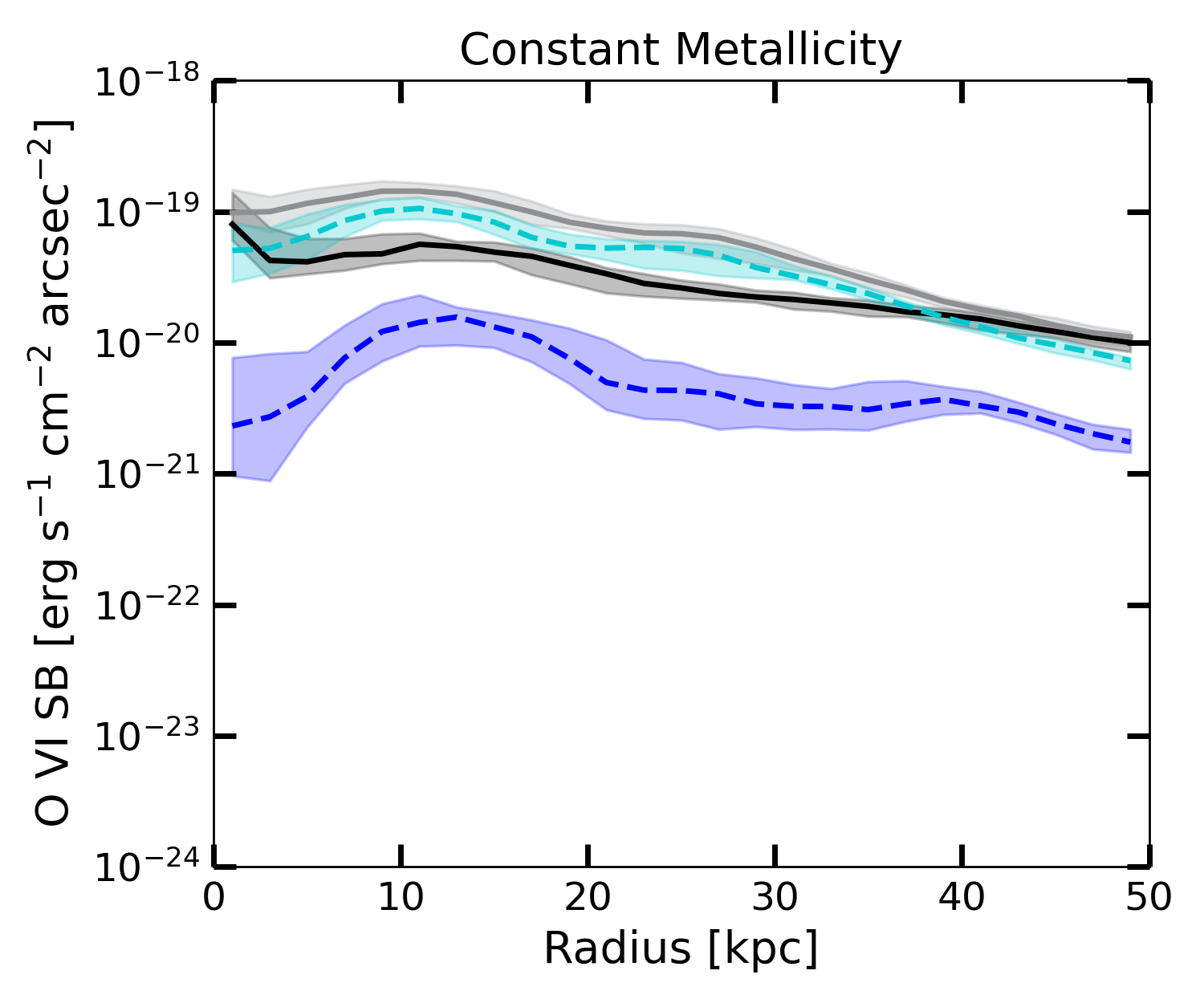}
    \caption{Left: \OVI\ SB profiles for the fiducial (black and blue) and no-feedback (gray and cyan) runs of Tempest for all gas (solid) and for inflowing gas (dashed), averaged across $z=0.1$--0, using the metallicity evolved within the simulation. The shaded regions show the 25--75\% spread of the surface brightness profiles over this time. The no-feedback run of Tempest has much lower \OVI\ SBs than the fiducial run, although the SB is still dominated by inflowing gas. Right: \OVI\ SB profiles in the fiducial and no-feedback runs of Tempest calculated assuming that all gas is held at a fixed metallicity of $Z_\odot$. Line colors and styles are the same as the left panel. The difference between the two runs decreases dramatically if they have the same metallicity, suggesting that the lack of feedback primarily affects the \OVI\ SB by not enriching the CGM with oxygen.}
    \label{fig:sb_no-feedback}
\end{figure*}

The left panel of Figure~\ref{fig:sb_no-feedback} shows the \OVI\ SB profile for the no-feedback run of Tempest compared to the fiducial run of Tempest, and again it is clear that the SB profile is much lower without feedback. The dashed curves show the \OVI\ SB profile from inflowing gas in both the no-feedback Tempest and fiducial Tempest, and in both cases the SB profile is dominated by the inflowing gas, as we saw before. We might even expect that the no-feedback run would have \emph{higher} \OVI\ SBs, because it is much more inflow-dominated than the fiducial run (and indeed the contribution of inflows to the SB profile are closer to the total SB profile in the no-feedback run), so why, then, does the lack of outflows so greatly suppress \OVI\ emission?

The answer lies in the metallicity: the right panel of Figure~\ref{fig:sb_no-feedback} shows the \OVI\ SB profiles again for both the fiducial and no-feedback Tempest runs, but now the emission is calculated assuming the metallicity is fixed at $Z_\odot$ for both simulations. The SBs are higher when the metallicity is fixed at a high value because there are more \OVI\ ions when there is generally more oxygen in the inner CGM. In addition, the difference in \OVI\ SBs between the fiducial and no-feedback cases is greatly reduced when both are held fixed at a constant, and the same, metallicity. This suggests that the drastic reduction by 1--2 dex in the \OVI\ SB for the no-feedback run of Tempest is actually driven by a lack of metals, and thus lack of \OVI\, in the inner CGM compared to the fiducial case \citep[a similar dependence of \OVI\ on black hole feedback-driven metallicity was found by][]{Sanchez2019}. Combined with Fig.~\ref{fig:sb_vs_Mh-den-temp-Z} that showed no dependence of median \OVI\ brightness with median CGM metallicity, it would appear that metallicity acts as a threshold rather than a scaler: once the CGM gas reaches a minimum metallicity that allows for \OVI\ to be prevalent (perhaps $\log (Z/Z_\odot)\sim-1.5$, the lowest values in Fig.~\ref{fig:sb_vs_Mh-den-temp-Z}), the brightness of the emission no longer strongly depends on the metallicity. It is also possible that the trend with metallicity is not purely due to enrichment, but is also related to cooling because cooling times are shorter when metals are present. We caution that simply forcing a constant solar metallicity without determining its effect on cooling times, as we did in Fig.~\ref{fig:sb_no-feedback} (right panel), will not fully correct for the lack of enrichment from turning off feedback.

Fig.~\ref{fig:no-feedback} (right panel) shows that the no-feedback rerun of Tempest exhibits a sudden drop in the \OVI\ SB at $\sim10$ Gyr ($z\sim0.35$), similar to the drop seen in Blizzard's \OVI\ SB time evolution (see Fig.~\ref{fig:sb_vs_time}). In Blizzard's case, the CGM metallicity dropped when the galaxy became nearly quiescent and feedback was no longer populating the CGM with metals. In the no-feedback Tempest rerun, feedback only populates the ISM with metals without launching them to the CGM, so this does not explain the sudden drop. Instead, it is the fact that the last (minor) merger occurs around 10 Gyr, so once the metal-enriched ISM material is fully accreted onto the main galaxy, there are few metals left outside the galaxy. This highlights that gas stripped from nearby satellites can contribute to observed CGM \OVI\ emission.

Although the outflows themselves are too hot to emit much in \OVI\ (Fig.~\ref{fig:gas-hists}), stellar feedback is important to populate the inner CGM with metals so that the warm and dense inflowing material is enriched enough to emit in \OVI. This suggests that the clumpy inflows that dominate the \OVI\ emission maps in the fiducial FOGGIE simulations contain significant recycled material that is enriched in oxygen. Without feedback, the inflows cannot be enriched, and would not be visible in \OVI\ emission.

%%%%%%%%%%%%%%%%%%%%%%%%%%%%%%%%%%%%%%%%%%%%%%%%%%%%%%%%%%%%%%%%%%%%%%%%%%%%%%%%%%%%%
\section{Discussion} \label{sec:discussion}

\subsection{\OVI\ emission as a tracer of the baryon cycle} \label{subsec:baryon_cycle}

In this work, we have seen that the brightest regions of \OVI\ emission originate in small structures (shells surrounding clumps) close to the galaxy, that the \OVI\ surface brightness is dominated by inflowing gas, and that stellar feedback is responsible for enriching this inflowing material in oxygen. Putting all this together suggests a picture in which \OVI\ emission traces an important piece of the baryon cycle: recycling of gas and metals back into the galaxy. \OVI\ cannot trace pristine inflows from the intergalactic medium through the CGM and onto the galaxy, because there is not enough oxygen in pristine or nearly pristine gas. It also cannot trace galactic wind fluid alone (within the assumptions of the stellar feedback scheme used in FOGGIE), because that material is too hot for \OVI\ ion fractions to be large. Instead, the \OVI\ emission maps are consistent with a picture where metal-enriched material flows out of the galaxy, mixes, cools, and/or condenses with material in the CGM, which then falls onto the galaxy in small, clumpy structures.

Such a ``galactic fountain" has been extensively studied in regards to the Milky Way's High Velocity Clouds (HVCs). Many of the observed HVCs in both \ion{H}{1} and more highly-ionized metal lines exhibit inflowing velocities that suggest they are participating in accretion of gas onto the Milky Way \citep{Bregman1980,Sembach2003,Fox2004,Lehner2012,Lehner2022a,Marasco2022}, and similar evidence for ionized gas accretion has been found in external galaxies as well \citep{Fraternali2008,Zheng2017}. Models and simulations have put forth a picture in which enriched outflows mix with cooler gas in the halo, creating intermediate-temperature gas in interface regions surrounding cooler clouds that can radiatively cool efficiently and rain back onto the galaxy \citep{Fraternali2013,Armillotta2016,Melso2019,Li2020}. Simulations have also found that the ionized gas, potentially including up to \OVI, could be dominating the accretion rate over \ion{H}{1} gas \citep{Joung2012,Lucchini2025} and that this gas is likely to have been previously enriched by cycling through the central galaxy or other nearby galaxies \citep{Hafen2019}. Using the FOGGIE simulations, \citet{Augustin2025} found that cooler clumps tend to be surrounded by warmer, more highly ionized shells. The strong dependence of emission on density means \OVI\ emission is more likely to pick up these warm shells surrounding dense, cold material than it is to reveal the diffuse warm phase, making it an excellent tracer of the gas phase that could be dominating galactic accretion.

\begin{figure*}
    \centering
    \includegraphics[width=0.8\linewidth]{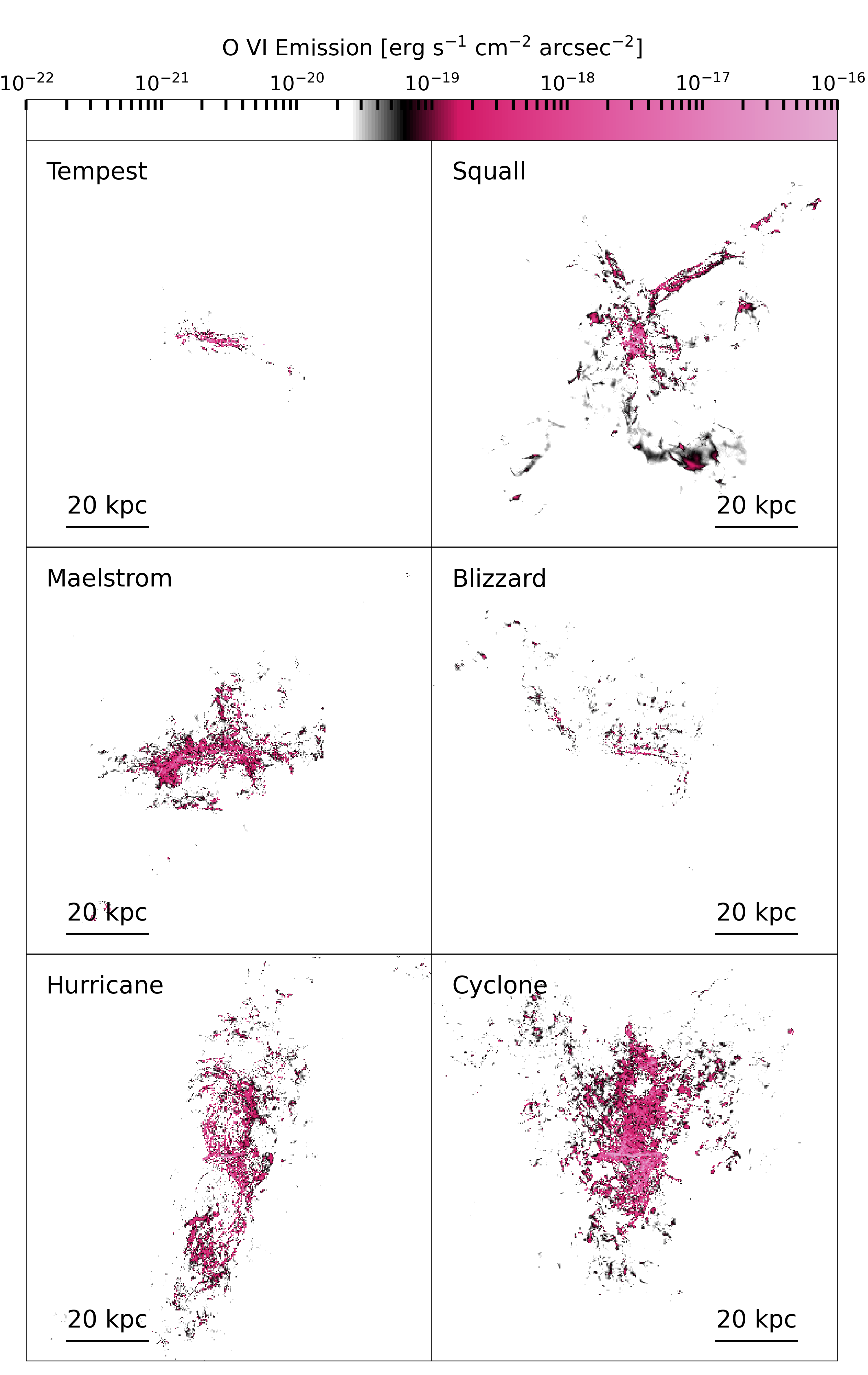}
    \caption{Surface brightness of \OVI\ emission from each of the six galaxies in the FOGGIE suite, each oriented edge-on and shown at $z=0$. Black on the color bar indicates a detection limit of $10^{-19}$ ergs s$^{-1}$ cm$^{-2}$ arcsec$^{-2}$ and red-pink indicates strength of detected emission, which shows that likely only the clumpy structures in the map are bright enough to be detected.} \label{fig:maps_z0_limit}
\end{figure*}

Studies of \OVI\ absorption have found a link between prevalence of CGM \OVI\ absorption and SFR of host galaxies \citep{Chen2009,Tumlinson2011,Tchernyshyov2023}, just as we find here with \OVI\ emission. A dependence of CGM \OVI\ column densities on SFR of host galaxies also manifests as the metallicity dependence we found in Section~\ref{subsec:feedback}: a lack of CGM enrichment would lead to both weaker \OVI\ emission and lower \OVI\ column densities. \citet{Sameer2024} discovered a similar observational finding: high \OVI\ column densities are primarily located in higher-metallicity gas. Taken altogether, the prevalence of \OVI\ in both CGM emission and absorption surrounding galaxies with higher star formation rates or more enriched gas suggests a strong link between \OVI\ and galaxy evolution.

Observations with FUSE and HST have detected faint \OVI\ emission just beyond the disk of a handful of galaxies \citep{Hayes2016,Chung2021a} and located within strong galactic outflows \citep{Kim2024,Ha2025}. The detections close to galaxy disks may be tracing the same warm inflow processes we predict to be generating \OVI\ emission, but the \OVI\ emission in outflows seem to be tracing radiative cooling in the outflows themselves. The FOGGIE simulations exhibit very hot outflows that do not contribute to \OVI\ emission, but warm, ionized outflows that may glow in \OVI\ are detected for many galactic systems in absorption \citep[e.g.,][]{Heckman2015}. The picture of warm, ionized envelopes surrounding cooler clouds that we have found here for accretion could also be the correct picture for emission in outflowing gas. Recent simulation studies have shown that cool clouds can be entrained and swept up in hot outflows, and the interface regions between the cool clouds and surrounding hot gas are prone to radiative cooling that allows them to potentially be detected in emission \citep{Gronke2020,Tan2021}.

Whether it is tracing inflows or outflows, \OVI\ emission seems to require enriched material and potentially short-lived warm interfaces. It is clear that wherever \OVI\ may be detected in emission, the baryon cycle is likely at work.

\subsection{FOGGIE \OVI\ emission is a lower limit} \label{subsec:lower_limit}

The majority of observational information about \OVI\ in the CGM currently available comes from absorption spectroscopy, not emission. Here, we use the observed \OVI\ column densities in absorption to attempt to estimate if FOGGIE's predicted \OVI\ emission is consistent with observed \OVI\ \emph{absorption}. The FOGGIE galaxies have lower \OVI\ column densities, by a factor $\sim1$--2 dex, than the observed \OVI\ columns in external low-$z$ galaxies \citep{Tumlinson2011} and in M31 \citep{Lehner2020}. The KODIAQ-Z \citep{OMeara2015,OMeara2017,Lehner2022b} survey found \OVI\ in absorption at $z\approx2$--2.5, and \citet{Peeples2019} compared the \OVI\ column densities of an earlier version of the FOGGIE simulations at $z=2$ with these, finding that FOGGIE produces lower \OVI\ column densities by $\sim1$--2 dex than observed at high redshift as well.

Low \OVI\ column densities in FOGGIE could be due to lower CGM gas densities, lower CGM metallicities, or too hot or too cold CGM temperatures in the simulated galaxies compared to observed galaxies. Because the \OVI\ emission seems to arise primarily from shells surrounding small clumps of gas, it could also be that the simulated CGM is not clumpy enough, or the properties of the clumps (size, distribution, number) are different from those surrounding observed galaxies. \citet{Lehner2020} find that the FIRE-2 simulations tend to produce higher \OVI\ column densities in the CGM, especially when cosmic rays are modeled \citep{Ji2020}, but still lower than observed column densities by $\sim0.5$ dex. In FIRE-2, the \OVI\ appears to originate in a virialized halo heated by an accretion shock that slowly accretes onto the galaxy \citep{Hafen2019}, allowing time for mixing of metals into the halo. In FOGGIE, the enriched outflows are fast and hot and may not mix as thoroughly into the halo to produce large \OVI\ column densities.

Because the FOGGIE galaxies are selected to be fairly isolated by $z=0$, with quieter merger histories since $z=2$, it is possible that their CGM is under-enriched in metals as compared to galaxies in more populated or active environments. In addition, the lower simulation resolution in diffuse gas outside of the forced refinement region can lead to enhanced mixing and dilution of metals launched in the feedback of other galaxies \citep{Shah2025}. Together, these effects suggest a suppressed enrichment of intergalactic recycling, which \citet{Oppenheimer2012} showed can be a substantial source of metals at low-$z$. A lack of metals in the FOGGIE CGMs would reduce the \OVI\ column densities, and thus the predicted emission strengths. However, Fig.~\ref{fig:gas-hists} shows the CGM metallicity distribution in FOGGIE is broad and extends from significantly super-solar down to 0.01 solar, peaking near 0.1 solar. \citet{Prochaska2017} and \citet{Lehner2013} both find CGM metallicities approximately one-third solar are common in absorption surveys, so the metal enrichment of the FOGGIE CGMs does not appear to be lacking.

Quasar absorption line surveys obtain a one-dimensional skewer of information through any given galaxy's CGM, so it is difficult to determine the 3D distribution of gas to inform how the simulations may be mismatched with observations. \citet{Piacitelli2022} estimates the \OVI\ SB from the \OVI\ column density measurements of \citet{Werk2014} and indeed finds SB values 1--2 dex higher than our \OVI\ SB profiles, in line with the simple expectation that \OVI\ emission scales with \OVI\ absorption. Rather than attempt to correct the unknown factors that could contribute to the \OVI\ column density mismatch between the FOGGIE simulations and observations, we emphasize that the emission maps at both high and low redshift are very likely lower limits, possibly by an order of magnitude or more, and stress that future work forward-modeling simulations into emission mock observations is crucial.

\subsection{Other simulated \OVI\ emission predictions} \label{subsec:comparison}

Other studies have made predictions for emission maps of various ions using a variety of simulations. \citet{Corlies2016} predicted the likelihood of detecting emission from \OVI, among a handful of other metal ions, from cosmological zoom simulations. They found, as we do here, that \OVI\ emission is most likely to be detected close to the galaxy, within $\sim30$ kpc, in a clumpy or structured/filamentary morphology, and that the gas predominantly producing the \OVI\ emission is hot and dense. \citet{Augustin2019} similarly found that \OVI\ emission is clumpy and near the galaxy for galaxies of a similar mass range as we investigate here.

\citet{Corlies2020} made emission map predictions for the FOGGIE galaxy Tempest at $z=3$, finding similar results that \OVI\ emission is primarily structured and close to the center of the galaxy, as well as generally arising from the densest tail of the gas density distribution within a temperature range of $T\sim10^5$--$10^6$ K. The similarity of their findings at $z=3$ to the findings presented here at $0<z<1$ for the full FOGGIE suite suggests that the properties of \OVI\ emission are relatively consistent across a wide range in redshift within the FOGGIE simulations, as we find in Fig.~\ref{fig:sb_vs_time}. They also found that only a small fraction of the CGM had \OVI\ SBs above a detection threshold $\sim10^{-19}$ erg$^{-1}$ cm$^{-2}$ arcsec$^{-2}$, although note that the FOGGIE emission predictions are a lower limit (Sec.~\ref{subsec:lower_limit}).

\subsection{Predictions for \emph{Aspera} and other CGM emission probes} \label{subsec:Aspera}

\emph{Aspera}'s detection limit is expected to be $\sim1$--$5\times10^{-19}$ erg s$^{-1}$ cm$^{-2}$ arcsec$^{-2}$ \citep{Chung2021b}, so many of the dimmer \OVI\ structures predicted by the FOGGIE simulations would not be detected. Figure~\ref{fig:maps_z0_limit} shows the same \OVI\ emission maps for the same simulated galaxies as in Figs.~\ref{fig:maps_z0}, now with the color bar changed so that only emission above this limit is visible. The smooth, extended structures are far below the detection limit, leaving only the small, clumpy structures close to the galaxy disks to be detected in \OVI\ emission.

\begin{figure*}
    \centering
    \includegraphics[width=0.49\linewidth]{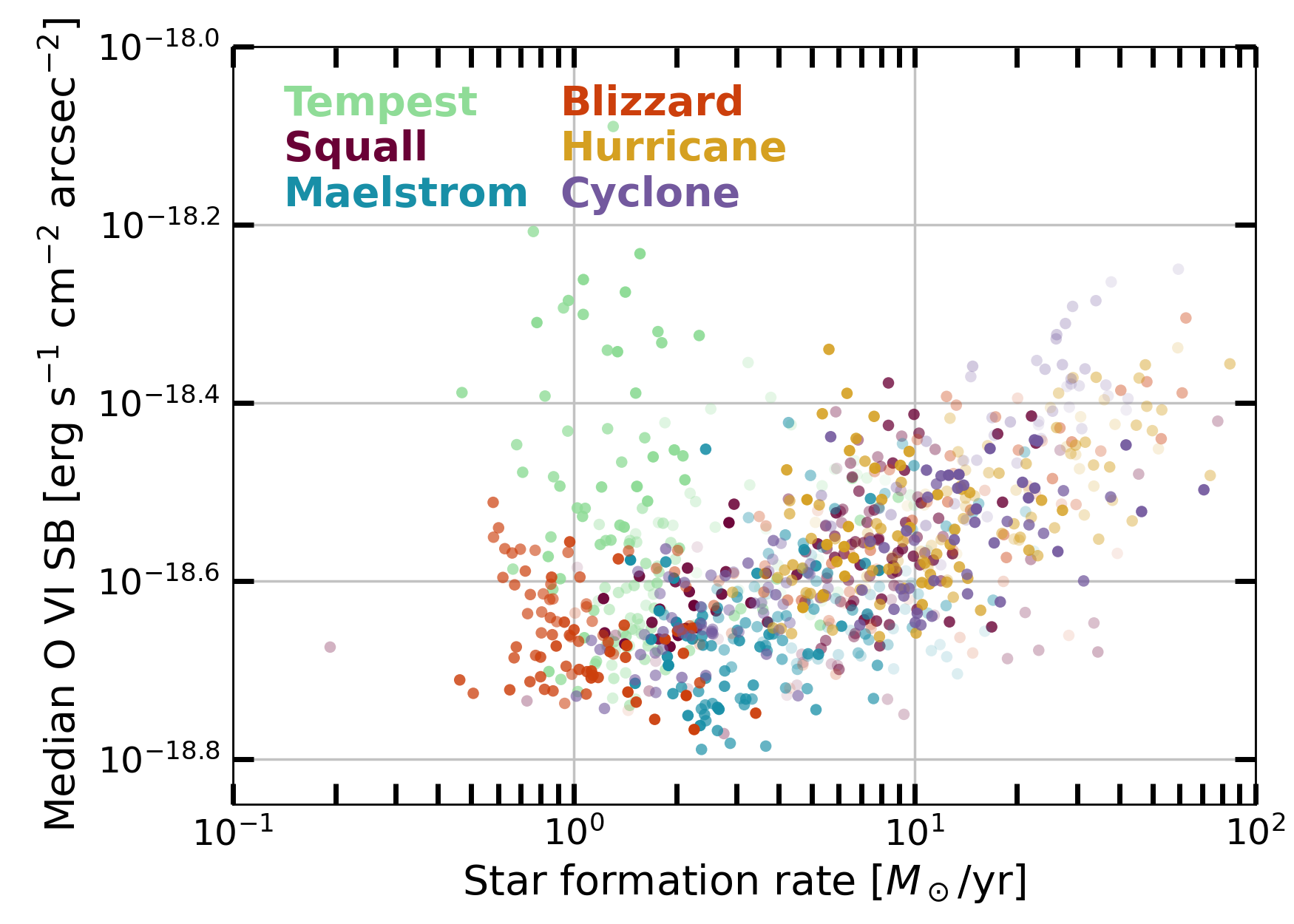}
    \includegraphics[width=0.49\linewidth]{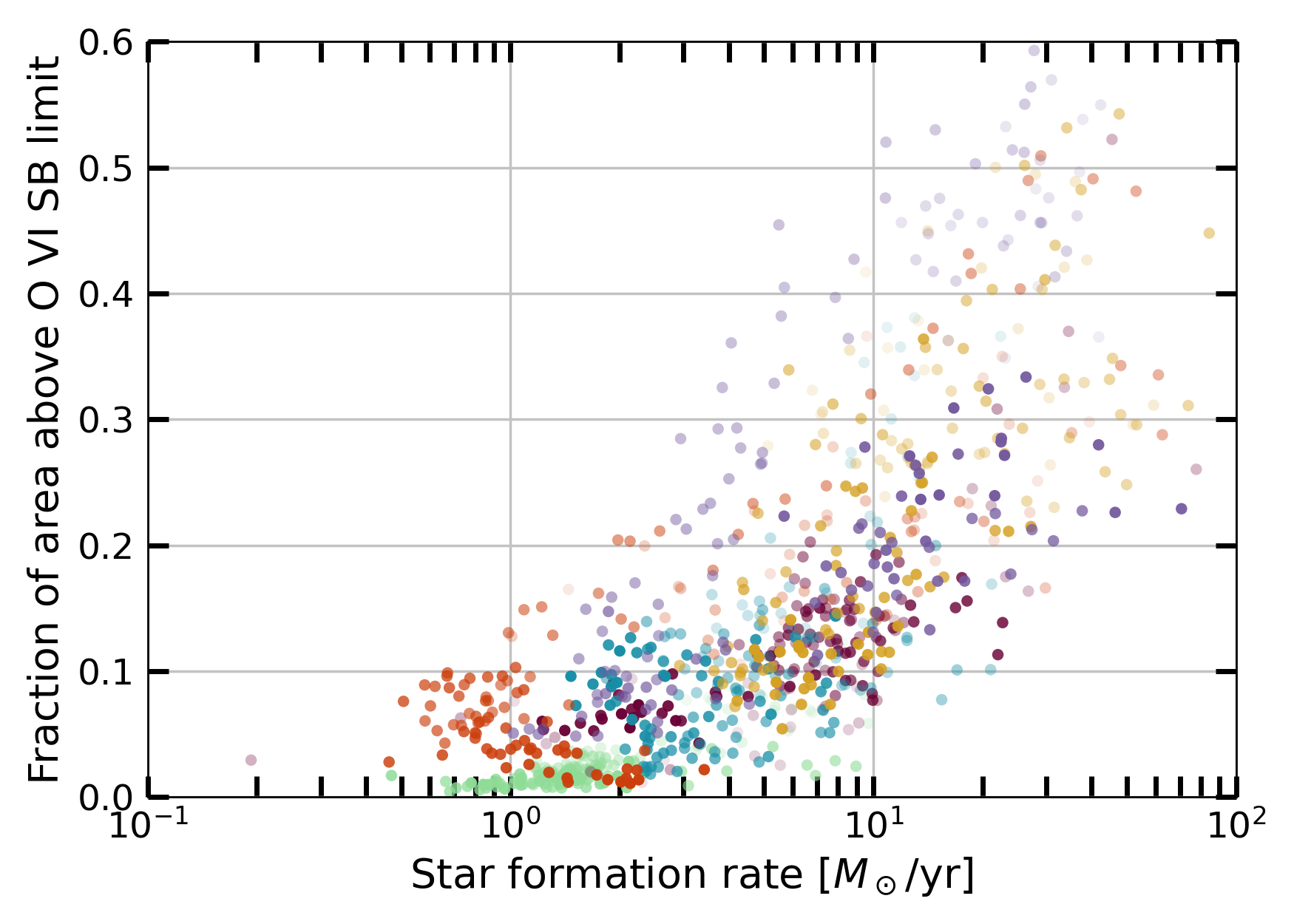}
    \caption{Left: Median \OVI\ SB within 20 kpc of the galaxy center vs. SFR of the central galaxy, for snapshots with $0<z<1$ for each of the six FOGGIE galaxies, calculated using only those pixels with SB above a detection limit of $10^{-19}$ ergs s$^{-1}$ cm$^{-2}$ arcsec$^{-2}$. The trend of increasing \OVI\ SB with increasing SFR is weaker, but still visible. Right: The fraction of the area within 20 kpc of the galaxy center that has an \OVI\ SB above \emph{Aspera}'s detection limit. The fractional area increases as the SFR increases, suggesting that bursts in SFR lead to both stronger and more widespread \OVI\ emission.}
    \label{fig:sb_vs_sfr_limit}
\end{figure*}

Even though there are not many pixels above the detection limit, the correlation of increasing median \OVI\ SB with increasing SFR still holds when considering only those pixels above \emph{Aspera}'s limit. The left panel of Figure~\ref{fig:sb_vs_sfr_limit} shows the median \OVI\ SB vs. SFR as in Fig.~\ref{fig:sb_vs_sfr}, but now the median \OVI\ is calculated from only those pixels that fall above \emph{Aspera}'s detection limit. The trend of increasing \OVI\ with increasing SFR is weaker, but it is still there---this suggests that \emph{Aspera} will be able to detect the baryon cycle as traced by \OVI. The dependence on SFR can be seen in a different way in the right panel of Figure~\ref{fig:sb_vs_sfr_limit}, which shows the fraction of the area within 20 kpc of the galaxy center with \OVI\ SB above \emph{Aspera}'s limit. The detectable area increases as the SFR increases, suggesting that bursts in the SFR not only create brighter \OVI\ emission close to the galaxy, but also create more widespread emission within a 20 kpc circle centered on the edge-on galaxy. Detecting this emission area requires CGM UV emission mappers to have a good enough spatial resolution to resolve the inner 20 kpc, but for galaxies near $z=0$, \emph{Aspera} and many other upcoming CGM UV emission probes will certainly meet this requirement. In addition, \citet{Saeedzadeh2025} shows that worse spatial resolution does not necessarily reduce the spatial area with detected emission above a given SB limit, just makes it more difficult to distinguish individual structures of emission.

Other proposed upcoming instruments may have deeper sensitivities than \emph{Aspera}. The proposed CubeSat, JUNIPER \citep{Witt2025}, and the proposed next-generation NASA flagship, Habitable Worlds Observatory, will both detect CGM UV emission to fainter values once built and launched \citep{Burchett2025}. \citet{Saeedzadeh2025} explores how much of the CGM can be detected and characterized by these and various other CGM emission instruments. Not only should CGM emission be readily detectable, but it also will unveil a new paradigm for our understanding of gas flows surrounding galaxies.

\subsection{Caveats} \label{subsec:caveats}

The FOGGIE simulations do not include modeling of a central supermassive black hole nor AGN feedback. \citet{Segers2017} and \citet{Oppenheimer2018} found that a fluctuating ionizing photon field from an AGN could drive \OVI\ out of ionization equilibrium and affect the column densities of \OVI\ in the CGM. AGN feedback can also push baryons out of the halo as well as provide an additional source of heating for the CGM. Section~\ref{subsec:foggie} shows that the lack of AGN in FOGGIE does not produce an over-massive CGM, and \citet{Silich2025} showed that the FOGGIE galaxies exhibit surprisingly similar mock X-ray properties to other simulations that do model AGN feedback, like IllustrisTNG. The hot gas in the FOGGIE CGM is a result of the stellar feedback model that deposits purely thermal energy, which drives hot and fast outflows that seem to mimic the results of AGN feedback in other simulations. The fraction of the CGM gas mass located in gas of $T\sim10^5$--$10^6$ K varies from 50\% to 70\% across the 6 FOGGIE halos, in line with estimates of \OVI-traced gas from COS-Halos \citep{Werk2014} and warm gas in EAGLE \citep{Schaye2015}, as collated by \citet{Tumlinson2017}. This suggests that the lack of AGN in FOGGIE is not producing an overly-cooled CGM with boosted \OVI. Future work modeling the impact of both stellar and AGN feedback on \OVI\ emission will be important to disentangle the effects of each and produce more accurate predictions.

We also note that the FOGGIE simulations do not include magnetic fields nor cosmic rays. Magnetic fields may affect the structure of the CGM, potentially collimating outflows and reducing mixing between phases \citep{vandeVoort2021}. Reducing mixing could also reduce the ability of outflows to enrich other CGM gas with metals, which would have a direct impact on the ability for non-outflowing gas to emit in \OVI\ (see Section~\ref{subsec:feedback}). Cosmic rays can provide non-thermal pressure support and reduce the quantity of shock-heated gas in the CGM, as well as launch much cooler and smoother outflows \citep{Butsky2018,Ji2020,Hopkins2021,Chan2022}. In some cases, this boosts the \OVI\ column densities out to larger radii from galaxies \citep{DeFelippis2024}, but the cooler nature of the CGM means this \OVI\ is likely photoionized and tracing more diffuse gas that may not produce strong emission. The picture we find of strong \OVI\ SB from warm envelopes surrounding cool clumps may not exist in a cosmic ray dominated CGM, if the entire CGM is cooler and more diffuse with less multiphase gas at high densities.

%%%%%%%%%%%%%%%%%%%%%%%%%%%%%%%%%%%%%%%%%%%%%%%%%%%%%%%%%%%%%%%%%%%%%%%%%%%%%%%%%%%%%
\section{Conclusions and Summary} \label{sec:conclusions}

In this work, we use the FOGGIE simulations to predict \OVI\ emission from the inner CGM of edge-on, Milky-Way-mass (at $z=0$) galaxies over the redshift range $z=1\rightarrow0$. Our main results are:

\begin{itemize}
    \item The brightest regions of \OVI\ emission arise from small structures that tend to have ``bubble" or ``shell" morphologies within 20 kpc of the center of galaxies (Fig.~\ref{fig:maps_z0}). There are significant variations in the structure of emission between the six FOGGIE galaxies studied here, but no preference in alignment with either the major or minor axis of the galactic disk (Fig.~\ref{fig:sb_axes_z0}).
    \item The gas that contributes most strongly to the emission does not simply trace the CGM structures with the most mass, but rather is primarily inflowing or not participating in fast flows, and is primarily warm with $T > 10^5$ K (Figs.~\ref{fig:sb_flows_z0},~\ref{fig:phase}, and~\ref{fig:sb_temp_z0}). The strongest \OVI\ emission originates in gas that is mostly in ionization equilibrium, but the edges of dense structures are most likely to be cooling too quickly for ionization equilibrium to be restored (Fig.~\ref{fig:tcool_teq}).
    \item \OVI\ surface brightnesses (SBs) in each pixel within 20 kpc of the center of each edge-on galaxy are broadly spread between $10^{-21}$ and $\sim10^{-16}$ erg s$^{-1}$ cm$^{-2}$ arcsec$^{-2}$, with median values decreasing from $\sim10^{-19.5}$ to $\sim10^{-20.5}$ erg s$^{-1}$ cm$^{-2}$ arcsec$^{-2}$ over the redshift range $z=1\rightarrow0$ (Fig.~\ref{fig:sb_vs_time}). When restricted to only those pixels above a surface brightness limit of $10^{-19}$ erg s$^{-1}$ cm$^{-2}$ arcsec$^{-2}$ (the expected sensitivity for the \emph{Aspera} SmallSat), the median surface brightness is constant over this redshift range.
    \item The median \OVI\ surface brightness weakly scales with increasing halo mass and increasing median density of the CGM within 20 kpc of the galaxy center, but has no correlation with median CGM temperature or metallicity within the same region (Fig.~\ref{fig:sb_vs_Mh-den-temp-Z}). The lack of correlation with CGM temperature is because \OVI\ ion fractions peak in the range $10^5$--$10^6$ K, so emission will primarily come from gas at this temperature regardless of the temperature distribution of gas in the inner CGM. Emission also comes primarily from the densest tail of the gas density distribution, but roughly follows the metallicity distribution of all gas in the inner CGM (Fig.~\ref{fig:gas-hists}).
    \item There is a strong positive trend between median \OVI\ SB in the inner CGM and the star formation rate (SFR) of the host galaxy (Fig.~\ref{fig:sb_vs_sfr}). Taken together with the finding that inflows dominate \OVI\ emission (Fig.~\ref{fig:sb_flows_z0}) and a no-feedback simulation shows nearly zero CGM \OVI\ emission (Fig.~\ref{fig:no-feedback}) unless the metallicity is artificially boosted (Fig.~\ref{fig:sb_no-feedback}), a picture emerges in which \OVI\ emission traces a galactic fountain where inflows are enriched in oxygen by outflows launched by stellar feedback. In FOGGIE, the outflows are too hot to contain \OVI\ themselves, but they enrich the CGM with metals. There is no obvious correlation between \OVI\ brightness and CGM metallicity, but there is a correlation in time between SFR and CGM metallicity for the strongest bursts and drops in the SFR (Fig.~\ref{fig:sb_vs_time}, bottom panels), suggesting that metal enhancement plays a role in \OVI\ emission but is not the only driver.
    \item For reasonable sensitivity limits for upcoming CGM emission probes, like the \emph{Aspera} SmallSat, \OVI\ emission is limited to only the brightest structures in the inner CGM (Fig.~\ref{fig:maps_z0_limit}). However, the FOGGIE simulations underestimate \OVI\ absorption column densities by $\sim1$--2 dex (Section~\ref{subsec:Aspera}), so the emission maps presented here are likely conservative lower limits. Even with this sensitivity limit, the strong correlation between \OVI\ SB and host galaxy SFR is detectable (Fig.~\ref{fig:sb_vs_sfr_limit}).
\end{itemize}

Taken altogether, our results highlight the power of \OVI\ emission as a probe of the baryon cycle, which plays an important role in galaxy evolution. \OVI\ SBs increase during starburst events, due to enrichment of the inner CGM with metals like oxygen. Gas mixing and recycling results in oxygen-enriched material falling back toward the galaxy in small, clumpy structures surrounded by bright \OVI-emitting shells, reminiscent of the high velocity clouds seen in the Milky Way halo. In this way, \OVI\ emission traces the sites of recycled accretion in galactic fountains that can dominate the total gas accretion onto Milky Way like galaxies.

Emission probes are a key part of the future for CGM studies, as they will allow us to map and disentangle the gas flows occurring between galaxies and their environments. UV instrument technology has advanced significantly to the point where great strides in CGM science can be accomplished by small satellites, and we are on the cusp of amazing discoveries that will be enabled by these new instruments.

\begin{acknowledgments}
The authors thank the anonymous referee for suggestions that improved the quality of the work. CL thanks the \emph{Aspera} Science Team for the impetus to carry out this investigation. CL also thanks Jake Bennett and Scott Lucchini for useful discussions in the CGM@CfA weekly chats.
Support for this work was provided by NASA through the NASA Hubble Fellowship grant \#HST-HF2-51538.001-A awarded by the Space Telescope Science Institute, which is operated by the Association of Universities for Research in Astronomy, Inc., for NASA, under contract NAS5-26555. 
VS was supported for this work in part by NASA via an Astrophysics Theory Program grant \#80NSSC24K0772, HST AR \#16151, and HST GO \#17093.
CT was supported for this work by NASA via a Theoretical and Computational Astrophysics Networks grant \#80NSSC21K1053 and JWST AR \#5486.
RA acknowledges funding from the European Research Council (ERC) under the European Union's Horizon 2020 research and innovation programme (grant agreement 101020943, SPECMAP-CGM).
AA acknowledges support from the INAF Large Grant 2022 ``Extragalactic Surveys with JWST" (PI Pentericci) and from the European Union - NextGenerationEU RFF M4C2 1.1 PRIN 2022 project 2022ZSL4BL INSIGHT.

Computations described in this work were performed using the publicly-available \textsc{Enzo} code (\href{http://enzo-project.org}{http://enzo-project.org}), which is the product of a collaborative effort of many independent scientists from numerous institutions around the world. Their commitment to open science has helped make this work possible.

Resources supporting this work were provided by the NASA High-End Computing (HEC) Program through the NASA Advanced Supercomputing (NAS) Division at Ames Research Center and were sponsored by NASA's Science Mission Directorate; we are grateful for the superb user-support provided by NAS.
\end{acknowledgments}

\software{\textsc{astropy} \citep{Astropy2013,Astropy2018},  
          CLOUDY \citep{Ferland2013},
          \textsc{matplotlib} \citep{matplotlib},
          \textsc{numpy} \citep{numpy},
          \textsc{scipy} \citep{scipy},
          \textsc{yt} \citep{yt}}

\begin{contribution}
    CL conceived the project and led the analysis and writing of the paper. MSP, BWO, and JT are the principal investigators of the FOGGIE collaboration, obtained funding for and developed and ran the simulations used in this paper. BWO also contributed to the interpretation of the results. LC wrote the code pipeline for obtaining emissivity from CLOUDY tables and VS ran the CLOUDY code to generate tables, and both also contributed their own experience with similar analysis to interpretation of results. NL, ACW, JKW, CWT, RA, and AA provided analysis suggestions and interpretation. BDS wrote the CIAOLoop code that was instrumental for generating the CLOUDY tables used, and also developed the initial conditions for the FOGGIE simulations. CJV is the PI of the Aspera instrument that is the inspiration for this work, and initiated the discussions that led to this project.
\end{contribution}

\appendix

\section{Time Dependence of Average CGM Properties} \label{app:time_depend}

\begin{figure}
    \centering
    \includegraphics[width=0.5\textwidth]{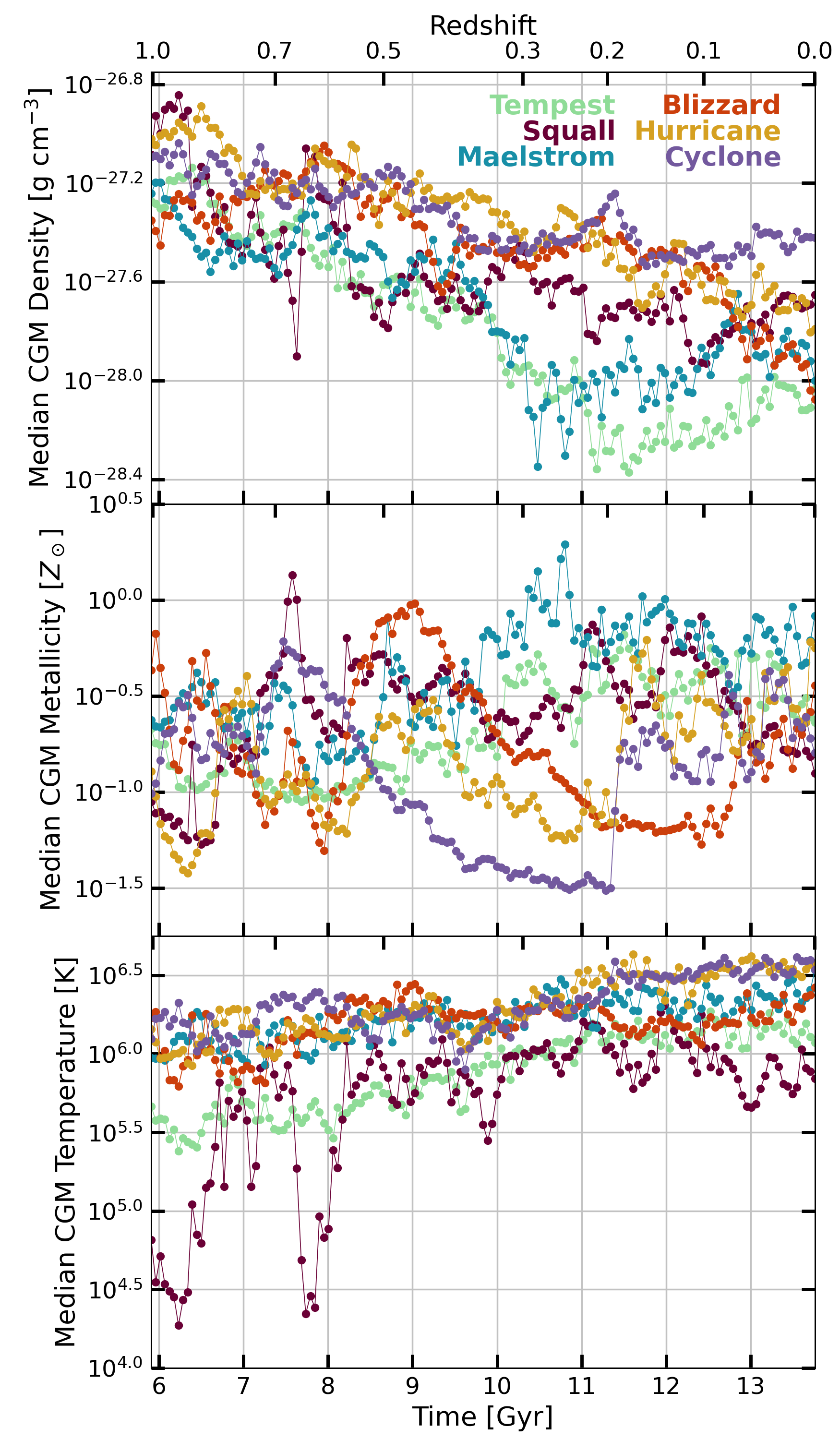}
    \caption{The volume-weighted median CGM density (top), metallicity (middle), and temperature (bottom) within 20 kpc of the galaxy center as a function a time, for the six FOGGIE galaxies. }
    \label{fig:prop_vs_t}
\end{figure}

To understand why the \OVI\ SB declines on average from $z=1\rightarrow0$, here we show the time evolution of the median CGM density, temperature, and metallicity within 20 kpc from the galaxy center. Figure~\ref{fig:prop_vs_t} shows the evolution over time of these gas properties for each of the six FOGGIE galaxies. The median CGM density declines by $\sim1$ dex over the redshift range from $z=1\rightarrow0$, possibly explaining the $\sim1$ dex decline in median \OVI\ SB over the same redshift range found in Fig.~\ref{fig:sb_vs_time} through the SB-density correlation shown in the top right panel of Figure~\ref{fig:sb_vs_Mh-den-temp-Z}. The average temperature of the inner CGM does not change much over the redshift range from $z=1\rightarrow0$, perhaps slightly increasing by $\sim0.3$ dex from $T\sim2.8\times10^6$ to $T\sim5.6\times10^6$ for five out of the six galaxies. There is significant scatter and no clear trend in the CGM metallicity over the redshift range from $z=1\rightarrow0$. Taken together with Fig.~\ref{fig:sb_vs_Mh-den-temp-Z}, these time evolutions suggest the shift in the median \OVI\ SB over the redshift range $z=1\rightarrow0$ is driven primarily by the shift in the median CGM density, as the median CGM temperature and metallicity do not vary as strongly.

\bibliography{mybib}{}
\bibliographystyle{aasjournalv7}

\end{document}